\documentclass[transmag]{IEEEtran}
\IEEEoverridecommandlockouts
\usepackage{url}
\usepackage{cite}
\usepackage{amsmath,amssymb,amsfonts}
\usepackage{graphicx}
\usepackage{textcomp}
\usepackage{xcolor}
\usepackage[linesnumbered,ruled,vlined]{algorithm2e}
\usepackage{lipsum}
\usepackage{mathtools}
\usepackage{cuted}
\usepackage{amsmath}
\usepackage{caption}
\usepackage{subcaption}
\usepackage{multirow}
\begin{document}
\title{Fairness Based Energy-Efficient 3D Path Planning of a Portable Access Point: A Deep Reinforcement Learning Approach \\
}
\author{Nithin Babu,~\IEEEmembership{Student Member,~ IEEE,}\,
       Igor Donevski,~\IEEEmembership{Student Member, ~IEEE,}\,\\
       Alvaro Valcarce,~\IEEEmembership{~Senior Member,~IEEE,}, Petar Popovski,\,\IEEEmembership{~Fellow, ~IEEE},\\ Jimmy Jessen Nielsen,~\IEEEmembership{~Member,~IEEE,}, and 
       Constantinos B. Papadias,\,\IEEEmembership{~Fellow,~IEEE.} 
\thanks{©2022 IEEE. Personal use of this material is permitted. Permission from IEEE must be obtained for all other uses, in any
current or future media, including reprinting/republishing this material for advertising or promotional purposes, creating new
collective works, for resale or redistribution to servers or lists, or reuse of any copyrighted component of this work in other
works.
This version of the work has been submitted in the IEEE OJCOMS. This work is supported by the project PAINLESS which has received funding from the European Union’s Horizon 2020 research and innovation programme under grant agreement No 812991.}
\thanks{N. Babu and C. B. Papadias are with Research, Technology and Innovation Network (RTIN), Alba, 
The American College of Greece, Greece (e-mail: nbabu@acg.edu, cpapdias@acg.edu).}
\thanks{N. Babu, I. Donevski, P. Popovski, J. J. Nielsen, and C. B. Papadias, and with Department of Electronic Systems, Aalborg University, Denmark (e-mail: \{niba, igordonevski, petarp, jjn, copa,\}@es.aau.dk).}
\thanks{A. Valcarce is with Nokia Bell-Labs France (e-mail: alvaro.valcarce\_rial@ nokia$-$bell$-$labs.com).}}

 \IEEEtitleabstractindextext{\begin{abstract}
 In this work, we optimize the 3D trajectory of an unmanned aerial vehicle (UAV)-based portable access point (PAP) that provides wireless services to a set of ground nodes (GNs).  Moreover, as per the Peukert effect, we consider pragmatic non-linear battery discharge for UAV’s battery. Thus, we formulate the problem in a novel manner that represents the maximization of a fairness-based energy efficiency metric and is named fair energy efficiency (FEE). The FEE metric defines a system that lays importance on both the per-user service fairness and the PAP’s energy efficiency. The formulated problem takes the form of a non-convex problem with non-tractable constraints. To obtain a solution we represent the problem as a Markov Decision Process (MDP) with continuous state and action spaces. Considering the complexity of the solution space, we use the twin delayed deep deterministic policy gradient (TD3) actor-critic deep reinforcement learning (DRL) framework to learn a policy that maximizes the FEE of the system. We perform two types of RL training to exhibit
the effectiveness of our approach: the first (offline) approach keeps the positions of the GNs the same throughout the training phase; the second approach generalizes the learned policy to any arrangement of GNs by changing the positions of GNs after each training episode. Numerical evaluations show that neglecting the Peukert effect overestimates the air-time of the PAP and can be addressed by optimally selecting the PAP's flying speed. Moreover, the user fairness, energy efficiency, and hence the FEE value of the system can be improved by efficiently moving the PAP above the GNs. \color{black}As such, we notice massive FEE improvements over baseline scenarios of up to 88.31\%, 272.34\%, and 318.13\% for suburban, urban, and dense urban environments, respectively\color{black}.
 \end{abstract}
 }
 \maketitle
\section{Introduction}
\color{black} To provide seamless network connectivity, it is expected that future radio access networks implement much denser deployments of small cells, that imply very high deployment costs. A more cost-efficient solution for serving a set of ground nodes (GNs) is to use an unmanned aerial vehicle (UAV) that carries a radio access node, hereafter referred to as a portable access point (PAP) \cite{survey1}.
\color{black} The third generation partnership project (3GPP) item \cite{3GPP} proposes the architecture and Quality-of-Service (QoS) requirements for such a system. The ability to have a controllable maneuver and the presence of line-of-sight (LoS) dominant air to ground channels \cite{cov1} make it appropriate for applications such as data collection from wireless sensor networks (WSNs), enhancing the cellular coverage, remote sensing, emergency deployments, and so on \cite{survey1}. The main drawback of the PAP system is its limited air-time which is a function of the capacity of the onboard battery unit and its power consumption profile. The air-time of a PAP is defined as the duration it remains aloft. The power consumed by a PAP varies with its mode of flying; for instance, a PAP consumes the maximum amount of power when it climbs vertically up, whereas the power consumption can be the least during a horizontal flight at an certain non-zero velocity \cite{filip}. Hence, the air-time of a PAP can be increased by suitably selecting its flying mode and velocity. Moreover, the available capacity of a PAP battery unit is a non-linear function of the power-draw profile of the PAP \cite{peukert}. Additionally, the air to ground channel LoS probability and the path loss between a PAP and a GN are proportional functions of the elevation angle and 3D distance between them, respectively \cite{cov1}. Consequently, the trajectory of a PAP can be used as a tool to increase its air-time and improve the channel to a GN. Hence, in this work, we design a 3D trajectory for a PAP that maximizes the number of bits transmitted per Joule of energy consumed while guaranteeing a fair service to the GNs measured in terms of fair energy efficiency (FEE) of the system.
\begin{figure}{}
\includegraphics[width=\columnwidth]{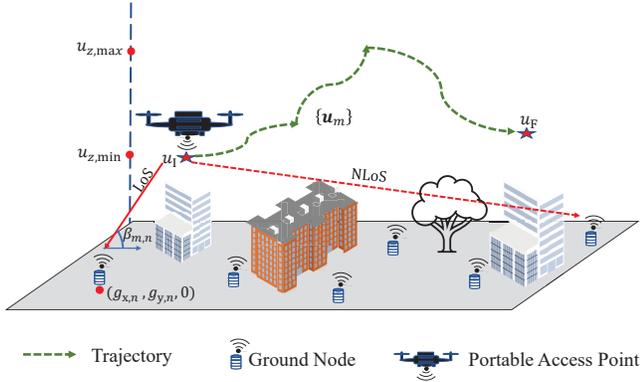}
\caption{Trajectory determination scenario.}\label{fig:system_model}
\end{figure}
\subsection{Related Works}
The works in \cite{cov1}-\cite{survey_energy} consider UAV placement optimization and trajectory design problems with main objectives as maximizing coverage area, throughput, air-time, energy efficiency, and minimizing mission time, power consumption, e.t.c. The authors of \cite{cov1}-\cite{cov5} consider the 3D placement of UAV(s) to maximize the coverage area. In \cite{cov1}, the authors propose a probabilistic LoS-non-LoS (NLoS) air to ground channel model and use it to find the optimal hovering altitude of a stationary UAV that maximizes the coverage area. \cite{cov2} and \cite{cov3} find the optimal altitude that maximizes the coverage area of a multi- and single-UAV system using circle packing theory. \cite{cov4} uses a combination of exhaustive search and maximal weighted area algorithm to propose an optimal UAV placement method that maximizes the number of users covered, whereas \cite{cov5} considers the placement optimization of a dynamic standalone drone equipped with a steerable antenna. The work in \cite{pow1} proposes a power-efficient deployment of multiple UAVs which are used as aerial base stations to collect data from ground Internet of Things (IoT) devices, whereas \cite{pow2} and \cite{pow3} consider minimizing the total transmit power of a drone base station by considering a downlink communication scenario.

The authors of \cite{rate1}-\cite{rev3} consider the average throughput of a UAV-based aerial communication system as the performance metric. \cite{rate1} and \cite{rate2} maximize the minimum average throughput by considering an uplink communication between a set of GNs and a UAV, whereas \cite{rate3}, \cite{rate4}, \cite{rev3}, and \cite{rev1}  consider a downlink communication scenario. In \cite{ee1}-\cite{babu3}, the authors consider the aerial vehicle's energy consumption while proposing an energy-efficient UAV(s) deployment policy. In \cite{ee1}, the authors propose a tractable power consumption model for a single-rotor rotary-wing UAV and use it to design a 2D trajectory that consumes the least amount of energy. In \cite{babu3}, we extend the model to a multi-rotor UAV and propose a 2D trajectory for a PAP that maximizes the number of bits transmitted per Joule of energy consumed while following a fly-hover-communicate protocol to serve the users. The algorithm given in \cite{ee2} designs an energy-efficient 2D trajectory for a fixed-wing UAV, while in \cite{babu2}, we determine a set of energy-efficient hovering points using circle packing theory. The works \cite{fairness1}-\cite{fairness3} use a deep reinforcement learning (DRL) technique to design a UAV(s) placement policy that guarantees fair service to the users. \cite{fairness1} uses the UAV trajectory as a tool to achieve fairness in terms of learning staleness, which reflects the learning time discrepancy among the users. The proposed policies of \cite{fairness2} and \cite{fairness3} achieve fairness in terms of coverage and throughput, respectively. Comprehensive lists of works that consider placement optimization of a UAV-based system are available in \cite{survey_path} and \cite{survey_energy}.
\subsection{Main Contributions and Paper Organization}
The works in \cite{cov1}-\cite{rate4} propose UAV trajectory design algorithms that either maximize communication-related parameters such as the coverage area and sum or average throughput or minimize the transmit power. The works mainly design a 2D trajectory or represent the 3D optimal UAV(s) positioning problems as two subproblems that optimize the vertical and horizontal positioning of the UAV(s) recurrently. Even though the problem formulations to maximize the energy-efficiency in \cite{ee1}-\cite{babu3} consider the aerial vehicle's power consumption, the solutions are again 2D flight trajectories. \color{black}
 Please note that for an energy-limited system such as a PAP, maximizing the number of bits transmitted per Joule of energy consumed while ensuring user fairness is paramount. Maximizing the throughput for a given energy budget is different from maximizing the energy efficiency since each movement of the UAV should maximize the throughput and minimizes the energy consumption simultaneously. Moreover, a UAV consumes different power during its axial climb and forward flight modes. Neglecting this, as in \cite{rev3}, \cite{rev1}, and \cite{fairness2}, falsely overestimates the air-time of a UAV resulting in the initiation of the early-landing procedure before completing the planned trajectory. Furthermore, \cite{rev3} and \cite{fairness2} propose trajectory planning and resource allocation schemes for high-mobility users in which the trajectory parameters and the resources are allocated to guarantee high instantaneous throughput fairness between all users. Even though the proposed fairness metric is ideal for analyzing the performance of the considered scenario, it might be sub-optimal for an IoT application such as data collection from an IoT network. For such applications, long-term fairness metrics are more suitable. For instance, consider a scenario in which the PAP is deployed to deliver a file of a given size to all the users by the end of the trajectory. In this case, the service fairness could be measured at the end of the trajectory; if all the users are delivered with an equal amount of bits on an average by the end of the trajectory, the fairness between the users will be high. 
 
 Suppose the PAP flies near to a user in a given time instant. In that case, it is more efficient to allocate more resources to that user since the communication channel to the user, as well as the throughput, will improve. However, to guarantee a high long-term user fairness, the later segments of the trajectory should be closer to the remaining users. Finally, none of the above works consider the Peukert effect seen in Li-ion batteries that are typically used in UAVs. Neglecting the Peukert effect overestimates the air-time of the PAP, resulting in initiating the early-landing procedure before completing the planned trajectory. In practise, the PAP will be flying at different velocities resulting in different power consumption; hence the remaining air-time of the PAP varies after each action as a non-linear function of the power consumption. This affects the system's energy efficiency since the number of trajectory segments varies as a non-linear function of the power consumption profile.
\color{black}
In essence, the 3D trajectory design of a PAP that maximizes the fairness-based energy efficiency while factoring in the UAV power consumption and the Peukert effect has, to the best of our knowledge, not yet been considered in the literature. The main contributions of this work are summarized as follows:
\begin{itemize}
    \item We propose a method to model the non-linear Peukert effect of the PAP battery using the data points from a data sheet;
    \item We additionally propose an algorithm to estimate the air-time of a PAP by considering the Peukert effect and the PAP power consumption profile;
    \item We introduce a user fairness-based energy efficiency metric called the  fair energy efficiency that considers user fairness, sum throughput, and the PAP propulsion power consumption;
    \item Finally, we implement a twin delayed deep deterministic policy gradient (TD3)- based 3D path planning algorithm to design a 3D trajectory for the PAP that maximizes the FEE value of the system.
\end{itemize}

This paper is structured as follows: Section \ref{system_model} explains the system setup, propagation environment, the 3D power consumption model of the PAP, and the FEE metric. In the section, we also detail the Peukert effect of the PAP battery and propose an algorithm to estimate the air-time of the PAP. Section \ref{DRL} includes the problem formulation to maximize the FEE of the system and the solving methodology. Section \ref{numerical_evaluation} presents the main findings through numerical evaluations, and elaborates the significance of the results. Finally, Section \ref{conclusion} summarizes the main findings of this work. All the quantities are in SI units unless otherwise specified.
\section{System Model}\label{system_model}
In this work, we consider a PAP deployed to serve a set of $N$ GNs. Each GN $n\in \mathcal{N}=\{1,2,3,..,N\}$ is located at $\mathbf{g}_{\mathrm{n}}=[\mathbf{g}_{\mathrm{h},{n}},0]$ of Cartesian space $(x,y,z)$ with $\mathbf{g}_{\mathrm{h},{n}}=[g_{\mathrm{x},{n}},g_{\mathrm{y},{n}}]$, as shown in Fig. \ref{fig:system_model}. The PAP flies along a 3D path to serve the set of GNs. Both the PAP and the GNs are assumed to be equipped with omni-directional antennas. 
\subsection{PAP Trajectory Model}
The optimal flying path of the PAP is obtained by dividing the total air time $T$ into $M$ time segments of length $\delta_t$ each such that $T=M\delta_t$ \cite{ee1}. The value of $\delta_t$ is chosen so that within each segment the PAP can be assumed to fly with a constant velocity, and the change in path loss values between the PAP and each GN is insignificant, i.e., $\delta_t v_{\text{max}}\leq \Delta$ where $v_{\text{max}}$ is the maximum speed of the UAV and $\Delta$ is the maximum change in distance below which the path loss values between the PAP and each GN remain stationary. Consequently, the path of the PAP can be represented using $M+1$ points, whose locations are denoted as $\mathbf{u}_{m}=[\mathbf{u}_{\mathrm{h},m},u_{\mathrm{z},{m}}]$, $m\in \mathcal{M}=\{1,2,3,..,M+1\}$ where $\mathbf{u}_{\mathrm{h},m}=[u_{\mathrm{x},{m}},u_{\mathrm{y},{m}}]$ is the projection of the PAP location on the horizontal plane. The length of each segment and the maximum PAP velocity are constrained as,
\begin{IEEEeqnarray}{rCl}
\|\mathbf{u}_{m+1}-\mathbf{u}_{m}\| &= & \delta_t v_{m} \leq \delta_t v_{\text{max}}\leq \Delta \,\,\forall m\in\mathcal{M}^{'}, 
\end{IEEEeqnarray}
where $\mathcal{M}^{'}=\{1,2,..M\}$. \color{black}In a particular segment, the PAP follows a time-division multiple access (TDMA) scheme to serve the GNs: let $T_{m,n}$ be the time allocated to the $n^{\text{th}}$ GN while the PAP is flying in the $m^{\text{th}}$ path segment with a speed of $v_m$ m/s, then,
\begin{IEEEeqnarray}{rl}
\Sigma_{n=1}^{N}T_{m,n} = \delta_t \quad \forall m\in \mathcal{M}^{'}. 
\end{IEEEeqnarray}
\color{black}
\subsection{Propagation Environment}
The communication channel between the PAP and a GN at a given time can be either LoS or NLoS depending on the relative position of the GN with respect to the PAP and the blockage profile of the environment.
The LoS and NLoS path loss values can be expressed as \cite{cov1}-\cite{pow1},
\begin{IEEEeqnarray}{rCl}
L^{\text{los}}_{m,n}& = & 
20\text{log}d^{\text{3D}}_{m,n}+20\text{log}f_c+20\text{log}\left(\dfrac{4\pi}{c}\right)+\eta^{\text{los}}\label{pathloss1}, \\
L^{\text{nlos}}_{m,n}& = & 
20\text{log}d^{\text{3D}}_{m,n}+20\text{log}f_c+20\text{log}\left(\dfrac{4\pi}{c}\right)+\eta^{\text{nlos}},\label{pathloss2}
\end{IEEEeqnarray}
with $d^{\text{3D}}_{m,n}=\sqrt{{d^{\text{2D}}_{m,n}}^{2}+u^{2}_{\mathrm{z},m}}$ and $d^{\text{2D}}_{m,n}=\|\mathbf{u}_{\mathrm{h},m}-\mathbf{g}_{\mathrm{h},n}\|$. $f_c$ and $c$ are the carrier frequency and the velocity of light, respectively. The corresponding probability of existence of a LoS link between the PAP and the $n^{\text{th}}$ GN while the PAP is in the $m^{\text{th}}$ path segment can be expressed as,
\begin{IEEEeqnarray}{rCl}
P^{\text{los}}_{m,n}& = & \dfrac{1}{1+a\exp{\left[-b(\beta_{m,n}-a)\right]}},
\end{IEEEeqnarray}
with $\beta_{m,n}=\text{arctand}\left(\dfrac{u_{\mathrm{z},m}}{d^{\text{2D}}_{m,n}}\right)$; $a$ and $b$  are the environment dependent parameters; $\eta^{\text{los}}$ and $\eta^{\text{nlos}}$ are the mean values of the respective additional path loss values due to long-term channel variations. For a given elevation angle, this additional path loss has a Gaussian distribution \cite{hourani}, and we use its mean value in this work\cite{cov1}-\cite{cov3}. The mean value depends on the building profile of the region and it is noticed that the change in the additional path loss within a particular propagation group (LoS/NLoS) is insignificant compared to the change in path loss value from one group to the other \cite{hourani},\cite{cov1}. This allows us to model the path loss with a constant gap between the two propagation groups as given in \eqref{pathloss1} and \eqref{pathloss2}. Hence, the expected spectral efficiency to the $n^{\text{th}}$ GN is given by,
\begin{IEEEeqnarray}{rCl}
\overline{R}_{m,n} &=&  P^{\text{los}}_{m,n}{R}_{m,n}^{\text{los}}+(1-P^{\text{los}}_{m,n}){R}_{m,n}^{\text{nlos}},\label{rmn}
\end{IEEEeqnarray}
where ${R}_{m,n}^{x}=\text{log}_2\left(1+\dfrac{P_t}{\sigma^210^{L^{x}_{m,n}/10}}\right)\,\,\forall x\in \{\text{los},\text{nlos}\}$;  
$P_t$ and $\sigma^2$ are the respective transmitted signal and noise power values.
\subsection{UAV power consumption model} \label{Uavpower}
\begin{table}[]
\caption{UAV's physical properties \cite{babu3}.}
\centering
\begin{tabular}{lll}
\hline
Label & Definition & Value \\ \hline
\hline
$W$ & Weight of the UAV in Newton & 24.5 N \\
$N_{\text{R}}$ & Number of rotors & 4 \\
$v_m$ &  UAV's horizontal flying velocity & -\\
$v_\text{tip}$ & Tip speed of the rotor & 102 m/s \\
$A_{\text{f}}$ & Fuselage area & 0.038 $\text{m}^2$  \\
$\rho(u_{\mathrm{z},m})$ & Air density at $u_{\mathrm{z},m}$  & - \\
$C_{D}$ & Drag Co-efficient & 0.9  \\
$A_{r}$ & Rotor disc area & 0.06 $\text{m}^2$ \\
$\Delta_p$ & Profile drag coefficient & 0.002 \\
$s$ & Rotor solidity & 0.05\\ \hline
\end{tabular}
\label{uavparameters}
\end{table}
\begin{figure}{h}
\centering
\includegraphics[width=0.5\columnwidth]{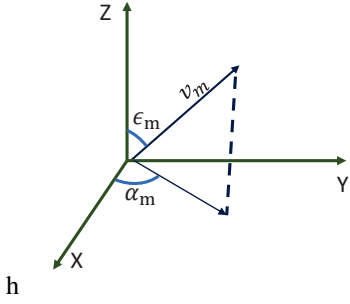}
\caption{Velocity Vector.}
\label{vel}
\end{figure}
 In this section, we provide the general expressions to calculate the total power consumed by the PAP during a considered time slot. The definitions and values\footnote[1]{\url{https://dl.djicdn.com/downloads/m100/M100_User_Manual_EN.pdf}} of all the variables used in this section are given in Table \ref{uavparameters}. In the $m^{\text{th}}$ time slot, the PAP moves from $\mathbf{u}_m$ to $\mathbf{u}_{m+1}$ in $\delta_t$ seconds. Then, as shown in Fig. \ref{vel}, the PAP velocity vector in the spherical coordinates system can be represented as $\mathbf{v}_m=(v_m,\alpha_m,\epsilon_m)$, in which $v_{m}=\|\mathbf{u}_{m+1}-\mathbf{u}_{m}\|/\delta_t$ is the speed of the PAP at which it travels from $\mathbf{u}_{m}$ to $\mathbf{u}_{m+1}$, $\alpha_m=\text{arctan}\left[({u}_{\mathrm{y},m+1}-{u}_{\mathrm{y},m})/({u}_{\mathrm{x},m+1}-{u}_{\mathrm{x},m})\right]$ and $\epsilon_m=\text{arctan}\left[\|\mathbf{u}_{\mathrm{h},m+1}-\mathbf{u}_{\mathrm{h},m}\|/({u}_{\mathrm{z},m+1}-{u}_{\mathrm{z},m})\right]$ are the azimuth and elevation angles of $\mathbf{u}_{m+1}$ with respect to the axes located at $\mathbf{u}_{m}$. In each time slot, the PAP is in one of the following flight conditions:
 \subsubsection{Forward flight ($v_m\neq 0,$ $ \epsilon_m\neq 0$)}
 The forward flight condition contains the following PAP flying modes: 1) the PAP moves along a plane that is parallel to the horizontal plane ($v_m\neq 0,$ $ \epsilon_m = 90^{\circ}$) commonly called as level forward flight; 2) forward (inclined) ascent or descent mode in which the PAP moves in the 3D space thereby changing all the 3 coordinates of its position ($v_m\neq 0,$ $ \epsilon_m \not\in\{ 90^{\circ}, 0^{\circ}$\}). The amount of power required to maintain this flight condition can be determined using \cite{babu3}, \cite{filip},  
 \begin{IEEEeqnarray}{l}\label{phfly}
P^{\text{fwd}}_{\text{uav}}(\mathbf{v}_{m})=\underbrace{ N_{\mathrm{R}}P_{\mathrm{b}}\left(1+\dfrac{3v^2_{m}}{v_{\text{tip}}^2}\right)}_{P_{\text{blade}}}+\underbrace{\dfrac{1}{2}C_{D}A_{\text{f}}\rho(u_{\mathrm{z},m})v^3_{m}}_{P_{\text{fuselage}}}\nonumber\\
+\underbrace{ W\left[\sqrt{\left(\sqrt{\dfrac{W^2}{4 N_{\mathrm{R}}^2 \rho^2 (u_{\mathrm{z},m})A_{\mathrm{r}}^2}+\dfrac{v_{m}^4}{4}}-\dfrac{v_{m}^2}{2}\right)}+\text{cos}\epsilon_m\right]}_{P_{\text{induce}}}\nonumber\\
\end{IEEEeqnarray}
where $P_{\text{b}}=\dfrac{\Delta_p}{8}\rho(u_{\mathrm{z},m}) s A_{\mathrm{r}} v^3_{\text{tip}}$ and $\rho(u_{\mathrm{z},m})=(1-2.2558.10^{-5} u_{\mathrm{z},m})^{4.2577}$. $P_{\text{blade}}$ and $P_{\text{fuselage}}$ are the powers required to overcome the profile drag forces of the rotor blades and the fuselage of the aerial vehicle that oppose its forward movement, respectively, while $P_{\text{induce}}$ represents the induced power from the rotation of rotors.
 \subsubsection{Hover ($v_m=0$)} In this mode, the PAP is static and its position is the same as that in the previous time slot. From \cite{babu3}, the hovering power consumption of a PAP is estimated using, 
 \begin{IEEEeqnarray}{rCl}\label{phover}
P^{\text{hov}}_{\text{uav}}&=& N_{\mathrm{R}}P_{\mathrm{b}}+ \dfrac{W^{3/2}}{\sqrt{4 N_{\mathrm{R}} \rho (u_{\mathrm{z},m})A_{\mathrm{r}}}}.
\end{IEEEeqnarray}
\subsubsection{Axial climb or descent ($v_m\neq 0,$ $ \epsilon_m=0$)}
Here, the PAP moves along the $+/-$ z-direction. Using (12.35) of \cite{filip}, the power required by the PAP to climb vertically ($\epsilon_m = 0$) is expressed as,
\begin{IEEEeqnarray}{rCl}
     P^{\text{vert}}_{\text{uav}} (v_{m})&=& \dfrac{W}{2}\left(v_{m}+\sqrt{v_{m}^2+\dfrac{2W}{N_{\mathrm{R}}\rho(u_{\mathrm{z},m})A_\mathrm{r}}}\right)+N_{\mathrm{R}}P_{\mathrm{b}}.\nonumber\\
 \end{IEEEeqnarray}
Hence, the total power consumed by the PAP while it flies along the $m^{\text{th}}$ path segment is calculated as,
\begin{IEEEeqnarray}{rCl}\label{puav}
P_{\text{uav}} (\mathbf{v}_{m})& = & \left\{\begin{matrix}
P^{\text{fwd}}_{\text{uav}}(\mathbf{v}_{m})&\quad  \text {if}\,\, v_m \neq 0 \,\& \,\epsilon_m \neq 0,\\
P^{\text{hov}}_{\text{uav}} & \quad  \text {if}\,\,v_m = 0,\\
P^{\text{vert}}_{\text{uav}}(v_{m})&\quad  \text {if}\,\, v_m \neq 0 \,\& \,\epsilon_m = 0.
\end{matrix}\right.
\end{IEEEeqnarray}
\subsection{Fair Energy Efficiency}\label{FEE_}
The fair energy efficiency (bits/Joule) of the system is expressed as,
\begin{IEEEeqnarray}{rCl}
\color{black}\text{FEE}\left(\mathcal{V}_M\right)&=&\dfrac{\color{black}\text{FI}\left(\mathcal{V}_M\right)\sum_{m=1}^{M}\sum_{n=1}^{N} D_{m,n}\left(\mathbf{v}_m\right)}{\sum_{m=1}^{M}\delta_tP_{\text{uav}}(\mathbf{v}_m)},\nonumber\\\label{eq_fee}
 \end{IEEEeqnarray}
 where $\color{black}D_{m,n}\left(\mathbf{v}_m\right)=BT_{m,n}\overline{R}_{m,n}$ is the number of bits transmitted to the $n^{\text{th}}$ GN while the PAP is in the $m^{\text{th}}$ segment, and
 \begin{IEEEeqnarray}{rCl}
\color{black}{\text{FI}}\left(\mathcal{V}_M\right)&=&\dfrac{\left[\sum_{n=1}^{N}\color{black} \overline{\text{D}}_{n}\left(\mathcal{V}_M\right)\right]^{2}}{N\sum_{n=1}^{N}\color{black}\overline{\text{D}}_{n}^2\left(\mathcal{V}_M\right)},
\end{IEEEeqnarray}
is the fairness index with  $\color{black}\overline{\text{D}}_{n}\left(\mathcal{V}_M\right)=\sum_{m=1}^{M}\text{D}_{m,n}\left(\mathbf{v}_m\right)/M$ giving the average number of bits transmitted to the $n^{\text{th}}$ GN by the end of the trajectory.  $\color{black}\text{FI}\left(\mathcal{V}_M\right)=1$ means the PAP sends equal number of data bits to the GNs when it completes the trajectory. $B$ is the total available bandwidth; $\mathcal{V}_M =\{\mathbf{v}_m,\forall m\in \mathcal{M}^{'}\}$. Considering the energy efficiency metric alone could allow the PAP to fly above a sub-set of GNs to maximize the energy efficiency by increasing the sum rate. Furthermore, the fairness index can be maximized either by maximizing the average number of bits transmitted to each GN or by minimizing it. The FEE metric defined in \eqref{eq_fee} is a weighted energy efficiency metric, where the weight is the fairness index. This forces the PAP to follow a 3D trajectory that maximizes energy efficiency and per-user fairness.
\subsection{The Peukert Effect}\label{peukert_effect}
 \begin{figure}{}
 \includegraphics[width=0.9\linewidth]{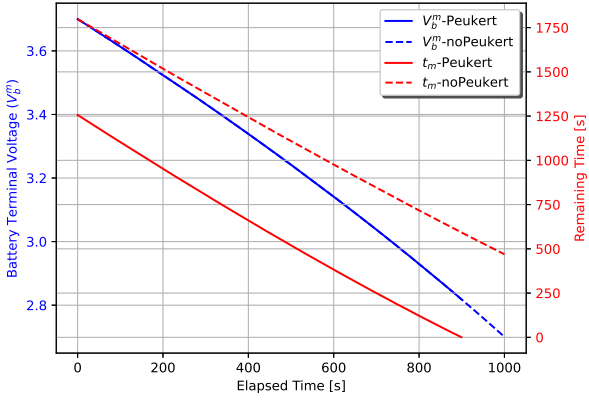}
 \caption{\color{black}Air-time with and without considering the Peukert effect for constant 200W power-draw until the battery discharges completely.}\label{peukert_fig}
 \end{figure}
A usual approach to estimate the maximum air-time of a PAP is to find the ratio of the initial onboard energy to the sum of instantaneous power consumption values \cite{ee1}-\cite{babu3}. This calculation has the fundamental assumption that the available discharge time of the PAP battery remains the same irrespective of the power-draw profile of the PAP. But, in practice, the battery discharge rate affects its available discharge time as shown in Fig. \ref{peukert_fig}, called the Peukert effect \cite{peukert}.

 Let $c_o$ be the rated capacity of a cell of the PAP battery unit in ampere-hours (Ah) and $t_o$ be the rated discharge time in hours (h). This means, if the PAP draws 1A of current from the cell, the cell will be completely discharged after $t_o$ hours. However, in practice, the current drawn by the PAP changes with time as a function of the power required and the terminal voltage of the battery:
 \begin{IEEEeqnarray}{rC}\label{pow_volt}
P_{\text{uav}}(\mathbf{v}_{m}) & = i^\mathrm{b}_m\cdot n^\mathrm{b}\cdot V^\mathrm{b}_m \,\,\,\,\forall m\in \mathcal{M}^{'},
\end{IEEEeqnarray}
where $n^\mathrm{b}$ is the number of battery cells connected in series to form the battery unit of the PAP with $V^\mathrm{b}_m$, the terminal voltage of a battery cell at the beginning of the $m^{\text{th}}$ time slot; also, $V^\mathrm{b}_{1}=V_o$ is the nominal voltage of the battery. Hence, the current drawn by the PAP during the $m^{\text{th}}$ slot is $i^\mathrm{b}_m=P_{\text{uav}}(\mathbf{v}_{m})/(n^\mathrm{b} \cdot V^\mathrm{b}_m)$. After the $m^{\text{th}}$ slot, the battery terminal voltage drops according to,
\begin{IEEEeqnarray}{rC}
V^\mathrm{b}_{m+1} &= V^\mathrm{b}_{m}-s_m^\mathrm{b}(i^\mathrm{b}_{m})\cdot i^\mathrm{b}_{m}\delta_t\,\,\forall m\in \mathcal{M}^{'},\label{vbm}
\end{IEEEeqnarray}
 where $s^{\mathrm{b}}_m(i^\mathrm{b}_{m})$ is the rate of change of terminal voltage per Ah that changes as a function of $i^\mathrm{b}_{m}$. In addition to the drop in the battery terminal voltage, the remaining discharge time also changes after each time slot according to,
 \begin{IEEEeqnarray}{rC}
t^\mathrm{b}_{m+1} & = t^\mathrm{b}_m \left(\dfrac{c^\mathrm{b}_m-i^\mathrm{b}_m\delta_t}{i^{\mathrm{b}}_{m+1}t^{\mathrm{b}}_m}\right)^{p^\mathrm{b}}\quad\,\,\forall m\in \mathcal{M}^{'},
\label{tm1}
\end{IEEEeqnarray}
with 
\begin{IEEEeqnarray}{rC}
t^\mathrm{b}_{1} & = t_o \left(\dfrac{c_o}{i^{\mathrm{b}}_{1}t_o}\right)^{p^\mathrm{b}};
\label{t1}
\end{IEEEeqnarray}
 $c_m^\mathrm{b}=t^\mathrm{b}_mi_m^\mathrm{b}$ with $c_1^\mathrm{b}=c_o$; $p^\mathrm{b}>1$ is the Peukert coefficient that depends on the type of the battery used; $i^{\mathrm{b}}_{m+1}$ is determined by substituting the voltage determined using \eqref{vbm} in \eqref{pow_volt} to guarantee a power output of $P_{\text{uav}}(\mathbf{v}_{m+1})$. 
The PAP should reach the destination either before the value of the terminal voltage reaches $V_{\text{cutoff}}$: $V^\mathrm{b}_{M+1} \geq V_{\text{cutoff}}$ or $t^\mathrm{b}_{M+1} \geq 0$. 
\begin{figure}{}
\includegraphics[width=0.9\linewidth]{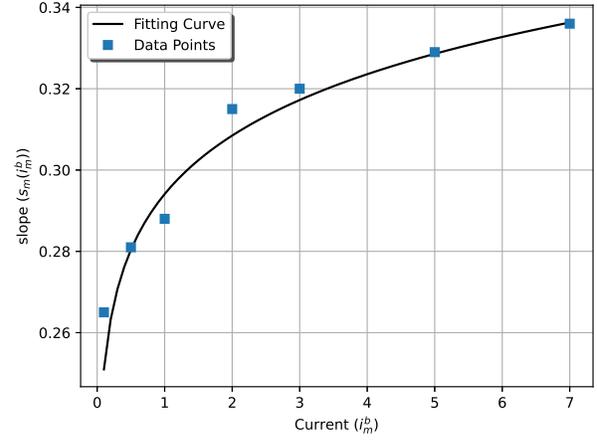}
\caption{\color{black}Variation of battery discharge slope for different discharge current.}\label{sbm_fitting}
\end{figure}

The slope, $s_m^\mathrm{b}(i^\mathrm{b}_{m})$ depends on the type of the battery used in the PAP. For a Li-ion battery with $c_o=4.5$ Ah, $t_o=3$ h, $i_o=1$ A, $V_{\text{cutoff}}=2.5$ V, and $V_o=3.7$ V, we perform a curve fitting over the variation of $s_m^\mathrm{b}(i^\mathrm{b}_{m})$ with regards to $i^\mathrm{b}_{m}$ using the data points from the data sheet \cite{datasheet} as shown in Fig. \ref{sbm_fitting}: 
\begin{IEEEeqnarray}{rC}\label{sbm}
s_m^\mathrm{b}(i^\mathrm{b}_{m}) &= f^\mathrm{b}_1\cdot {i^\mathrm{b}_{m}}^{f^\mathrm{b}_2},
\end{IEEEeqnarray}
where $f^\mathrm{b}_1 =   0.2941$, and  $f^\mathrm{b}_2 = 0.06888$. 

\color{black}The Peukert effect is better explained in Fig. \ref{peukert_fig}. The figure shows the voltage drop and the remaining discharge time of a typical Li-ion battery (commonly used in UAVs) during discharge when the PAP draws a power of 200 W continuously. As shown in the figure, a battery is useful until the terminal voltage or the remaining discharge time becomes lower than the corresponding threshold values (2.9 V and 30 Seconds, respectively), whichever happens first. The air-time of the PAP is defined as the time elapsed from the beginning till the battery is useful. As seen in the figure, neglecting the Peukert effect overestimates the air-time of the PAP. Thus a trajectory planned considering the availability of air-time determined without considering the Peukert effect will initiate the early-landing procedure before completing the trajectory. This affects the system's service fairness, sum rate, and energy efficiency. \color{black}
The PAP air-time for a given power profile considering the Peukert effect can be estimated using Algorithm 1.
\begin{algorithm}[]
\caption{PAP Air-Time Estimation}
Initialize $m=1$ $P_{\text{uav}}(\mathbf{v}_m)$, $t_o$, $V_{o}$, $c_o$, $V_{\text{cutoff}}$, $\delta_ t$;\\
 \While{1}
 {
    \If{($m==1$)}
    {
        $V_m^\mathrm{b}=V_{o}$; find $i_m^\mathrm{b}$ from \eqref{pow_volt}; find $t_m^\mathrm{b}$ using \eqref{t1};\\
    }
    \If{($t_{m}^\mathrm{b}\leq 0$)}
    {
    break;
    }
    find $s_m^\mathrm{b}(i_m^\mathrm{b})$ using \eqref{sbm};\\
    find $V^\mathrm{b}_{m+1}$ using \eqref{vbm};\\
    \If{($V^\mathrm{b}_{m+1}<V_{\text{cutoff}}$)}
    {
        break;\\
    }
    $m=m+1$; update $P_{\text{uav}}(\mathbf{v}_m)$;\\
    find $i_m^\mathrm{b}$ from \eqref{pow_volt}; $t_m^\mathrm{b}$ using \eqref{tm1};\\
 }
 \textbf{Output}:\,Air-Time: {$T_{\text{uav}} = m\delta_t$.}
 \label{endurance}
\end{algorithm}
\section{Trajectory Optimization Using DRL Method}\label{DRL}
In this section, we formulate the problem and use the deep reinforcement learning technique to design an optimal trajectory for the PAP that maximizes the FEE of the system.
\subsection{Problem Formulation}\label{problem_formulation}
The FEE of the considered system can be increased by suitably designing the 3D trajectory of the PAP. The corresponding problem can be formulated as,
\begin{IEEEeqnarray}{rCl}
\text{(P1)} & : & \underset{\color{black}\mathcal{V}_M}{\text{maximize}}\,\,\,\, {\color{black}\text{FEE}}\color{black}\left(\mathcal{V}_M\right), \nonumber\\
\text{s.t.}& & V^\mathrm{b}_{M+1} \geq V_{\text{cutoff}};  t_{M+1}^\mathrm{b} \geq 0,   \,\,\,\,\label{c1}\\
&& u_{\mathrm{x},m+1} = u_{\mathrm{x},m}+\underbrace{\delta_t v_{m} \text{sin}\,\epsilon_m \text{cos}\,\alpha_m}_{\Delta^\mathrm{x}_{{m}}(\mathbf{v}_{m})} \, \forall m\in \mathcal{M}^{'},\label{c5}
\end{IEEEeqnarray}
\begin{IEEEeqnarray}{rCl}
&& u_{\mathrm{y},m+1} = u_{\mathrm{y},m}+\underbrace{\delta_t v_{m} \text{sin}\,\epsilon_m \text{sin}\,\alpha_m }_{\Delta^\mathrm{y}_{{m}}(\mathbf{v}_{m})}\, \forall m \in \mathcal{M}^{'},\label{c6}\\
&& u_{\mathrm{z},m+1} = u_{\mathrm{z},m}+ \underbrace{\delta_t v_{m} \text{cos}\,\epsilon_m}_{\Delta^\mathrm{z}_{{m}}(\mathbf{v}_{m})} \forall m \in \mathcal{M}^{'}\label{c7},\\
&& u_{\mathrm{z},\text{min}} \leq u_{\mathrm{z},m} \leq u_{\mathrm{z},\text{max}}\quad \forall m\in \mathcal{M},\label{c10}\\
&& \Sigma_{n=1}^{N}T_{m,n} = \delta_t \quad \forall m \in \mathcal{M}^{'}, \label{c11}\\
& & \mathbf{u}_{M+1} = \mathbf{u}_{\mathrm{F}}; \mathbf{u}_{1} = \mathbf{u}_{\mathrm{I}},\label{c12}\\
& & \quad T_{m,n} \geq 0 \quad  \forall m\in \mathcal{M}^{'},n\in \mathcal{N}.\label{c13}
\end{IEEEeqnarray}
The objective function of (P1) maximizes the FEE; \eqref{c1} ensures that the PAP will not run out of onboard available battery capacity at any point of the trajectory. The x, y, and z coordinates of the PAP position are changed according to $\eqref{c5}-\eqref{c7}$, respectively. The flying region of the PAP is limited in the z-direction using $\eqref{c10}$ with $u_{\mathrm{z},\text{min}}$ and $u_{\mathrm{z},\text{max}}$ as the respective minimum and maximum permitted flying altitudes. \eqref{c11} is the TDMA scheduling constraint. \eqref{c12} constrains the initial and final positions of the PAP to be $\mathbf{u}_{\mathrm{I}}$ and $\mathbf{u}_{\mathrm{F}}$, respectively. (P1) is a non-convex optimization problem with a large number of optimization variables restricting the use of conventional convex optimization methods such as sequential convex programming\cite{scp}. Consequently, (P1) is equivalently represented as a Markov Decision Process (MDP) with continuous state and action spaces, and a DRL-based algorithm is proposed to design a 3D trajectory for the PAP that maximizes the FEE of the system.

The PAP is considered as an agent of the DRL framework; the framework takes the state observed by the PAP, $s_m$, and outputs an action, $a_m$. The agent receives a reward $r_m$ after taking the action $a_m$ that moves it from state $s_m$ to state $s_{m+1}$. The whole trajectory of the PAP is considered as an episode of the DRL framework; \color{black}an episode ends (i.e., $m=M+1$) if it runs out of the onboard battery capacity. It should be noted that the value of $M$ is not constant here, and it varies according to the profile of the PAP power consumption. Additionally, to model the FEE solely as a function of the PAP trajectory, we schedule the data transmission to each GN for a time that is proportional to the respective expected spectral efficiency (i.e. $T_{m,n}=\overline{R}_{m,n}/\sum_{n=1}^{N}\overline{R}_{m,n}$, $\forall m\in \mathcal{M}^{'}, n\in \mathcal{N}$). \color{black}
\subsection{PAP Trajectory as an MDP}
Since the next state and action of the PAP depend only on the present state of the PAP, we use a standard MDP representation as a 4-tuple $(\mathcal{S},\mathcal{A},\mathcal{P},\mathcal{R})$ with sets: state space $\mathcal{S}$, action space $\mathcal{A}$, probability of transition $\mathcal{P}$, and a state-action reward map $\mathcal{S} \times \mathcal{A} -> \mathcal{R}$.
\subsubsection{State Space, $\mathcal{S}=\{s_m\}$}
The state of the PAP consists of \color{black}3D-\color{black}coordinates of the PAP location written relative to the destination, the PAP's battery terminal voltage, the total energy consumed, coordinates of the GNs written relative to the horizontal projection of the PAP position, and the number of bits transmitted to each GN until the end of the $m^{\text{th}}$ time slot:
\begin{IEEEeqnarray}{rCl}
s_m&=&\{\{(\mathbf{u}_{m}-\mathbf{u}_{\mathrm{F}}), {V}^{\mathrm{b}}_m, E_{m}, \{(\mathbf{g}_{\mathrm{h},n}-\mathbf{u}_{\mathrm{h},m})\}, \{{D}^{\text{sum}}_{m,n}\}\},\nonumber\\ 
\end{IEEEeqnarray}
where ${D}^{\text{sum}}_{m,n}= \sum_{j=1}^{m}D_{j,n}(\mathbf{v}_j,T_{jn})$ is the total number of bits transmitted to the $n^{\text{th}}$ GN till the end of the $m^{\text{th}}$ time slot; $E_m=\sum_{j=1}^{m}\delta_tP_{\text{uav}}(\mathbf{v}_j)$ is the total energy consumed until the end of the $m^{\text{th}}$ time slot. Hence, $s_m$ has $5+3\cdot N$ dimensions. 
\subsubsection{Action Space $\mathcal{A}=\{a_m\}$}
Since all the state dimensions are functions of the 3D movement of the PAP, the action $a_m$ taken by the PAP is velocity-steered and can be expressed as a vector of dimension $3$: $a_m=\{c_{m}^\mathrm{x}, c_{m}^\mathrm{y},c_{m}^\mathrm{z}\}\in \left[-1,1\right]$ such that the components of the velocity vector are given by,
\begin{IEEEeqnarray}{rCl}
v_{m} &=& \sqrt{{c_{m}^\mathrm{x}}^2+{c_{m}^\mathrm{y}}^2+{c_{m}^\mathrm{z}}^2} \cdot \dfrac{v_{\text{max}}}{3},\label{vmm}\\
\epsilon_m &=& \text{arctan}\left(\dfrac{\sqrt{{c_{m}^\mathrm{x}}^2+{c_{m}^\mathrm{y}}^2}}{c_{m}^\mathrm{z}}\right),\\
\alpha_m &=& \text{arctan}\left(\dfrac{{{c_{m}^\mathrm{y}}}}{c_{m}^\mathrm{x}}\right)\label{alphamm}.
\end{IEEEeqnarray}
Moreover, if the action takes the PAP out of the altitude boundaries, the z-coordinate of the next state is readjusted to the corresponding boundary value.
\subsubsection{Reward Space $\mathcal{R}=\{r_m\}$}
The reward function determines how fast the PAP finds the optimal trajectory. Here, the primary objectives are to maximize the FEE and let the PAP reach the specified destination before the battery becomes obsolete by satisfying all the constraints of (P1). To efficiently map the above objectives, we leverage the reward shaping technique \cite{reward}.
Hence, the reward $r_m$ is expressed as,
\begin{IEEEeqnarray}{rCl}
r_m= f_m+p_m,
\end{IEEEeqnarray}
where,
\begin{IEEEeqnarray}{rCl}
p_m & = & \left\{\begin{matrix}
\color{black}\text{FEE}\color{black}(\mathcal{V}_{m+1}) & \text{if}\,\,  \text{FEE improves},\\
0 & \text{otherwise}/ \text{if}\,$m=M+1$,
\end{matrix}\right.
\label{positionreward}
\end{IEEEeqnarray}
is the position reward that encourages the PAP to move in a direction that improves the FEE of the system, and  
\begin{IEEEeqnarray}{rCl}
f_m & = & \left\{\begin{matrix}
\kappa_\mathrm{f}\cdot\color{black}\text{FEE} \color{black}(\mathcal{V}_{M}) & \text{if} \quad m = M+1,\\
0 & \text{else},
\end{matrix}\right.
\label{terminalreward}
\end{IEEEeqnarray}
is the terminal reward. The value of $\kappa_\mathrm{f}$ should be selected in a way that ensures the sum of the position rewards is always less than or equal to the terminal reward. $\kappa_\mathrm{f}$ is needed to balance the position and terminal rewards. Otherwise, the position reward would dominate over the terminal reward.

Since the defined MDP is deterministic, no randomness is considered and all transitions follow the agent’s decisions \cite{fairness1}. Therefore, the next state is a direct consequence of the current action of the agent.
\subsubsection{Episode Termination}
The FEE of the system increases when the PAP spends the maximum time over the air to serve the GNs. Hence, an episode is terminated when the remaining air-time ($T_{\text{uav}}$) of the PAP is equal to the minimum time required by the PAP to reach the destination from the current position with a speed of $v_{\text{max}}$: $T_{\text{uav},m}^{\text{min}} = \|\mathbf{u}_{m}-\mathbf{u}_\mathrm{F}\|/v_{\text{max}}$. The remaining air-time of the PAP can be estimated using Algorithm 1. Consequently, if an action takes the PAP to $\mathbf{u}_{m+1}$ and if $T_{\text{uav}}< T_{\text{uav},m+1}^{\text{min}}$, the action is discarded and the PAP moves to the destination from $\mathbf{u}_{m}$ with a speed of $v_{\text{max}}$ m/s. Accordingly, the GNs are served for a maximum amount of time while ensuring a safe landing of the PAP at the destination.   

\color{black}
The safety check explained above satisfies \eqref{c1} by ensuring sufficient energy available at the PAP to fly back to the destination after each action. The 3D coordinates of the PAP after taking an action are determined by substituting \eqref{vmm}-\eqref{alphamm} in \eqref{c5}-\eqref{c7}, satisfying the PAP movement constraints. The altitude constraint \eqref{c10} is satisfied by limiting the action space if such an action violates the constraint. The proposed heuristic time allocation in which the data transmission to each GN is scheduled for a time proportional to the respective expected spectral efficiency satisfies the TDMA constraints \eqref{c11} and \eqref{c13}. Finally, all the episodes start and end at $\mathbf{u}_{\mathrm{I}}$ and $\mathbf{u}_{\mathrm{F}}$, respectively  satisfying constraint \eqref{c12}.
\color{black}
\subsection{TD3-Based PAP 3D Path Design}\label{ee_pathdesign}
Here, the objective is to find the optimal policy $\pi$ that takes the current state of the PAP (agent) and gives an action that maximizes the expected return: $R_m=\sum_{i=m}^{M}\gamma^{i-m}r_m$, where $\gamma$ is a discount factor determining the priority of short-term rewards. The action value of a state, $Q_{\pi}(s_m,a_m)$, gives the expected return for starting in state $s_m$, taking action $a_m$, and then acting according to the policy $\pi$ forever after. The optimal action-value function is given by the Bellman equation as, 
\begin{IEEEeqnarray}{rCl}
Q^*(s_m,a_m) = \left[r_m + \gamma \max_{a_{m+1}} Q^*(s_{m+1}, a_{m+1})\right].\label{Qvalue}\nonumber\\
\end{IEEEeqnarray}
\begin{algorithm}[]\label{E2P2}
\caption{Energy Efficient 3D Path Planning}
Initialize the locations of GNs;\\
Initialize critic networks $Q_{\phi_1}$, $Q_{\phi_2}$, and actor network $\mu_\theta$ with random parameters $\phi_1$, $\phi_2$, $\theta$;\\
Initialize target networks $\phi_{1,\text{tgt}}\leftarrow \phi_{1}$, $\phi_{2,\text{tgt}}\leftarrow \phi_{2}$, $\theta_{\text{tgt}}\leftarrow \theta$\\
Initialize replay buffer $\mathcal{H}$;\\
\For{$\text{each episode}$}
{
Initialize the location of PAP to $\mathbf{u}_{I}$, $m=1$, $V_m^b=V_o$, $t_m=t_o$; ${d}=1$;\\
Formulate the state of the PAP $s_m$;\\
\While{the episode is not over}
{
The agent takes an action with exploration noise: $a_m=\mu_\theta(s_m)+\epsilon$, observe the reward $r_m$ and new state $s_{m+1}$;\\
store $(s_m,a_m,r_m,s_{m+1})$ in the replay buffer;\\
\If{replay buffer is sufficient}{
sample mini batch of $|\mathcal{H}^{'}|$  transitions from $\mathcal{H}$;\\
compute target actions using \eqref{exploration};\\
compute target using \eqref{target};\\
update critic networks by one step gradient descent using,
$\nabla_{\phi_j} \frac{1}{|\mathcal{H}^{'}|}\sum_{e_i\in \mathcal{H}^{'}}\Bigg( Q_{\phi_j}(s_i,\mu_{\theta}(s_i)) - y_i \Bigg)^2$ for $j \in \{1,2\}$;\\
\If{ it is time to update}{update policy network by one step gradient ascent using, $ \frac{\sum_{e_i\in \mathcal{H}^{'}} \nabla_{\mu_{\theta}(s_i)}Q_{\phi_1}(s_i,\mu_{\theta}(s_i))\nabla_{{\theta}}\mu_{\theta}(s_i)}{|\mathcal{H}^{'}|}$;\\
update target networks using,  $\theta_{\text{tgt}}=\tau \theta_{\text{tgt}}+(1-\tau \theta)$;\\
$\phi_{j,\text{tgt}}=\tau \phi_{j,\text{tgt}}+(1-\tau \phi_{j})$ for $j \in \{1,2\}$.}
}
$m=m+1$;\\
}
}
\end{algorithm}
In DRL, $Q^*(s_m,a_m)$ is approximated by a neural network $Q_{\phi}(s_m,a_m)$ with parameters $\phi$. Then the closeness of $Q_{\phi}(s_m,a_m)$ to $Q^*(s_m,a_m)$ is judged by evaluating the mean-squared Bellman error (MSBE) function:
\begin{IEEEeqnarray}{rCl}\label{MSBE}
L(\phi, {\mathcal H}) &=& \underset{\{e_i\} \sim {\mathcal H}}{{\mathrm E}}\left[
    \Bigg( Q_{\phi}(s_i,a_i) - y_i \Bigg)^2\right],
\end{IEEEeqnarray}
where 
\begin{IEEEeqnarray}{rCl}
y_i &=& r_i + \gamma (1 - d) \max_{a_{i+1}} Q_{\phi}(s_{i+1},a_{i+1}), \label{y_i}
\end{IEEEeqnarray}
is the target value and $d=1$ represents the terminal state. The expectation in \eqref{MSBE} is taken over a mini batch of experiences, $\{e_i\}=\{(s_i,a_i,r_i,s_{i+1},d)\}=\mathcal{H}^{'}$ sampled from the experience replay buffer, $\mathcal{H}$. The parameter $\phi$ is updated to minimize the MSBE. Since the considered action space is continuous, the evaluation of the MSBE is not trivial because of the $\max_{a_{i+1}} Q_{\phi}(s_{i+1},a_{i+1})$ term in the target value where the maximization has to be done over a continuous action space. To tackle this, we use an actor-critic framework-based twin delayed deep deterministic policy gradient (TD3) algorithm \cite{TD3}. An actor-critic framework uses an actor network that takes the state $s_m$ as input and outputs the action $a_m$, whereas the Q-value of the taken action $a_m$ at state $s_m$ is estimated by the critic network. At the end of the training, the actor network represents the optimal policy, $\pi$.  Hence, \eqref{MSBE} and \eqref{y_i} can be rewritten as,
\begin{IEEEeqnarray}{rCl}
L(\phi, {\mathcal H}) &=& \underset{\{e_i\} \sim {\mathcal H}}{{\mathrm E}}\left[
    \Bigg( Q_{\phi}(s_i,\mu_\theta(s_i)) - y_i \Bigg)^2\right]\label{MSBE_updated},\\ 
y_i &=& r_i + \gamma (1 - d)  Q_{\phi}(s_{i+1},\mu_\theta (s_{i+1})), \label{y_iupdated}
\end{IEEEeqnarray}
where $\mu_\theta$ is the actor network with parameters $\theta$ and $Q_{\phi}$ is the critic network with parameters $\phi$. From \eqref{MSBE_updated} and \eqref{y_iupdated}, the target $y_i$ depends on the same parameters we are trying to train: $\phi$ and $\theta$ which makes the MSBE minimization unstable. The solution is to use target networks that have sets of parameters which come close to $\phi$ and $\theta$, but with a time delay. The parameters of the target network are denoted as $\phi_{\text{tgt}}$ and $\theta_{\text{tgt}}$, respectively. In order to avoid the overestimation problem of the deep deterministic policy gradient (DDPG) algorithm \cite{DDPG}, \color{black}the TD3 algorithm proposed in \cite{TD3} uses\color{black}:
\subsubsection{Clipped Double-Q Learning} in which two critic networks are used instead of one, and uses the smaller of the two Q-values to form the targets in the MSBE functions:
\begin{IEEEeqnarray}{rCl}
y_i &=& r_i + \gamma (1 - d) \min_{j=1,2} Q_{\phi_{j, \text{tgt}}}(s_{i+1},\mu_{\theta_{\text{tgt}}}(s_{i+1}) ),\label{target}
\end{IEEEeqnarray}
where $Q_{\phi_{j, \text{tgt}}}$ for $j\in\{1,2\}$ are the corresponding target critic networks. Both  networks are then trained to minimize this target:
\begin{IEEEeqnarray}{rCl}
L(\phi_1, {\mathcal H}) &=& \underset{\{e_i\} \sim {\mathcal H}}{{\mathrm E}}{
    \Bigg[ Q_{\phi_1}(s_i,\mu_{\theta}(s_i)) - y_i \Bigg]^2},\\
L(\phi_2, {\mathcal H}) &=& \underset{\{e_i\} \sim {\mathcal H}}{{\mathrm E}}{
    \Bigg[ Q_{\phi_2}(s_i,\mu_{\theta}(s_i)) - y_i \Bigg]^2},
\end{IEEEeqnarray}
that avoids the overestimation problem;
\subsubsection{Delayed Policy Updates} through which the TD3 updates the policy ($\mu_\theta$) and target networks less frequently than the critic networks (once every $K$ critic networks update);
 \subsubsection{Target Policy Smoothing} which adds a clipped noise on each dimension of the action produced by the target policy network. After adding the clipped noise, the target action is then clipped to lie in the valid action range: $ [a_{\text{min}}, a_{\text{max}}]$,
 \begin{IEEEeqnarray}{rCl}\label{exploration}
 \mu_{\theta_{\text{tgt}}}(s_{i+1}) = \text{clip}\left(\mu_{\theta_{\text{tgt}}}(s_{i+1}) + \text{clip}(\epsilon,-c,c), a_{\text{min}}, a_{\text{max}}\right),\nonumber\\ 
 \end{IEEEeqnarray}
 where $\epsilon \sim \mathcal{N}(0, \sigma)$ and $\text{clip}(x,a,b) = \text{max}(\text{min}(x,b),a)$. This avoids the problem of developing an incorrect sharp peak for some actions by the  Q-function approximator. The steps to design a fair energy-efficient 3D trajectory for the PAP using the TD3 framework are given in Algorithm 2.
\section{Numerical Evaluation}\label{numerical_evaluation}
\begin{table}[]
\centering
\caption{Testing environment settings}
\begin{tabular}{lll}
\hline
Label & Definition & Value \\ \hline
\hline
$f_c$ & Channel carrier frequency & 5.8 GHz \\
$c$ & Velocity of light & $3 \cdot 10^8$ m/s \\
$B$ & Channel bandwidth for each GN & 40 MHz \\
$N_0$ & Noise spectral power & -174 dBm/Hz \\
$u_{z,\text{min}}$ & PAP's minimum flying altitude & 20 m \\
$u_{z,\text{maz}}$ & PAP's maximum flying altitude & 100 m \\
$v_\text{max}$ & Maximum achievable PAP speed & 24 m/s \\
$\delta_t$ & Time discretization Interval & 1 s \\
$P_t$ & Transmission Power & 23 dBm \\ \hline
\end{tabular}
\label{table:setting}
\end{table}
In this section, we present our main findings obtained through numerical evaluations. The evaluations consider a square area of 1000 $\times$ 1000 meters with 16 GNs. The values of all the environment-related parameters are listed in Table \ref{table:setting} \cite{fairness1}. To the best of our knowledge, the same overall setting has not been considered in the literature yet, hence we are comparing our results with \color{black} the following two baseline scenarios:\\
\begin{enumerate}
    \item \underline{Baseline 1:} the first maneuver executed by the PAP is a diagonal climb from $\mathbf{u}_{\mathrm{I}}$ up to the center of the region with coordinates $(500,500,100)$; hovers there until it only has sufficient energy to reach its destination; and flies reclined to the end of its trajectory $\mathbf{u}_{\mathrm{F}}$;
    \item \color{black}\underline{Baseline 2:} the first maneuver executed by the PAP is a diagonal climb from $\mathbf{u}_{\mathrm{I}}$ to (200,200,100). It then continues through the shortest path between the locations of the GNs until it only has sufficient energy to reach its destination. We determine the shortest path using the well known travelling salesman algorithm.  
\end{enumerate} 
\color{black}The simulations are done considering 3 different environment scenarios namely, suburban, urban and dense urban with $(a,b,
\eta^{\text{LoS}},\eta^{\text{NLoS}})$ parameters $(4.88,0.43,0.2,24)$, $(9.61,0.16,1.2,23)$, and $(12.08,0.11,1.8,26)$, respectively \cite{cov1}.
\begin{figure}{}
\includegraphics[width=0.8\columnwidth]{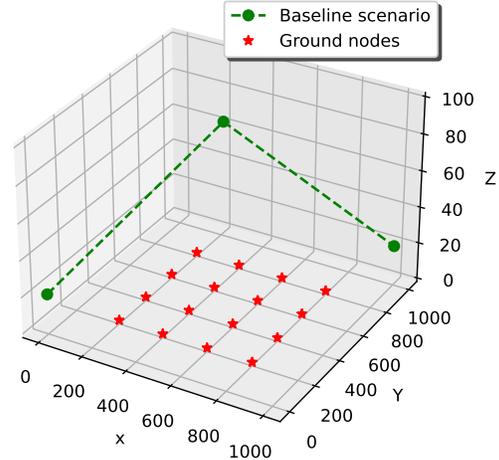}
\caption{Baseline1 scenario to compare the performance; the PAP starts at (0,0,20), flies to the center of the geographical region (500,500,100), hovers there until the battery capacity reaches the threshold value, flies back to the destination (1000,1000,20).}\label{baseline}
\end{figure}

\begin{figure}{}
\centering
\includegraphics[width=0.9\columnwidth]{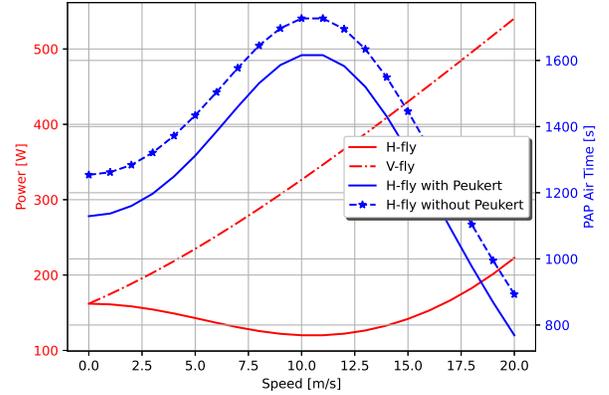}
\caption{\color{black}Variation of the PAP power consumption and air-time (endurance) as a function of the flying velocity. H-fly represents a level forward flight at the maximum height ($v_m\neq 0,$ $ \epsilon_m = 90^{\circ}$); v-fly represents axial climb or descent ($v_m\neq 0,$ $ \epsilon_m=0$).\color{black}}\label{pow_end}
\end{figure}

The architecture of actor and critic networks used in the simulations are listed in Table \ref{network_params}.
After an extensive experimentation, the values of various hyper parameters associated with the networks that give the maximum FEE value after training the networks for 1000 episodes are listed in Table \ref{simulation}.
\begin{table}[]
\caption{Network Architecture}
\centering
\begin{tabular}{llll}
\hline
Network(s) & Layer & Depth & Activation \\ \hline
\hline
Critic & Input Layer & 56 & $-$ \\
Critic & Hidden Layer 1 & 256 & ReLu \\
Critic & Hidden Layer 2 & 512 & ReLu \\ 
Critic & Hidden Layer 3 & 512 & ReLu \\
Critic & Output Layer & 1 & ReLu\\ \\
Actor & Input Layer & 53 & $-$\\
 Actor& Hidden Layer 1 & 256 & ReLu \\
 Actor& Hidden Layer 2 & 512 & ReLu\\ 
 Actor& Hidden Layer 3 & 512 & ReLu \\ 
 Actor& Output Layer & 3 & TanH\\
 \hline
\end{tabular}
\label{network_params}
\end{table}
\begin{table}[]
\centering
\caption{Network Parameters.}
\begin{tabular}{lll}
\hline
Label & Definition & Value \\ \hline
\hline
$\alpha$ & Actor learning rate& $ 10^{-4}$ \\
$\beta$ &  Critic learning rate & $ 10^{-3}$ \\
$|\mathcal{H}^{'}|$ &  Batch size & 64 \\
$|\mathcal{H}|$ & Replay buffer size & $2\times 10^5$ \\
$K$ & Network update interval & 2 \\
$\tau$ &Soft update factor &0.001\\
 $\gamma$ & Discount factor & 0.99 \\
 $\kappa_\mathrm{f}$ & -&1000\\\hline
\end{tabular}
\label{simulation}
\end{table}
Fig. \ref{pow_end} plots the PAP power consumption and air-time as a function of the speed. The vertical flying power consumption increases with speed since the PAP requires more power to overcome the downward drag force. When the PAP is flying horizontally, the power consumption initially decreases and then increases after 11 m/s: because the magnitude of power required to overcome the rotor-induced drag force decreases with the PAP velocity; in the low-speed regime, it dominates the power consumed to overcome the fuselage and rotor profile drag forces. Correspondingly, the maximum PAP air-time, using Algorithm 1, is obtained as 1616 seconds (s) when the PAP is flying at a speed of 11 m/s. The figure also shows the importance of considering the Peukert effect during the trajectory planning of the PAP; neglecting the Peukert effect overestimates the air-time of the PAP that could force the PAP to initiate landing procedure before completing the planned trajectory.

\begin{figure}
     \centering
     \begin{subfigure}[b]{0.49\columnwidth}
         \centering
         \includegraphics[width=\columnwidth]{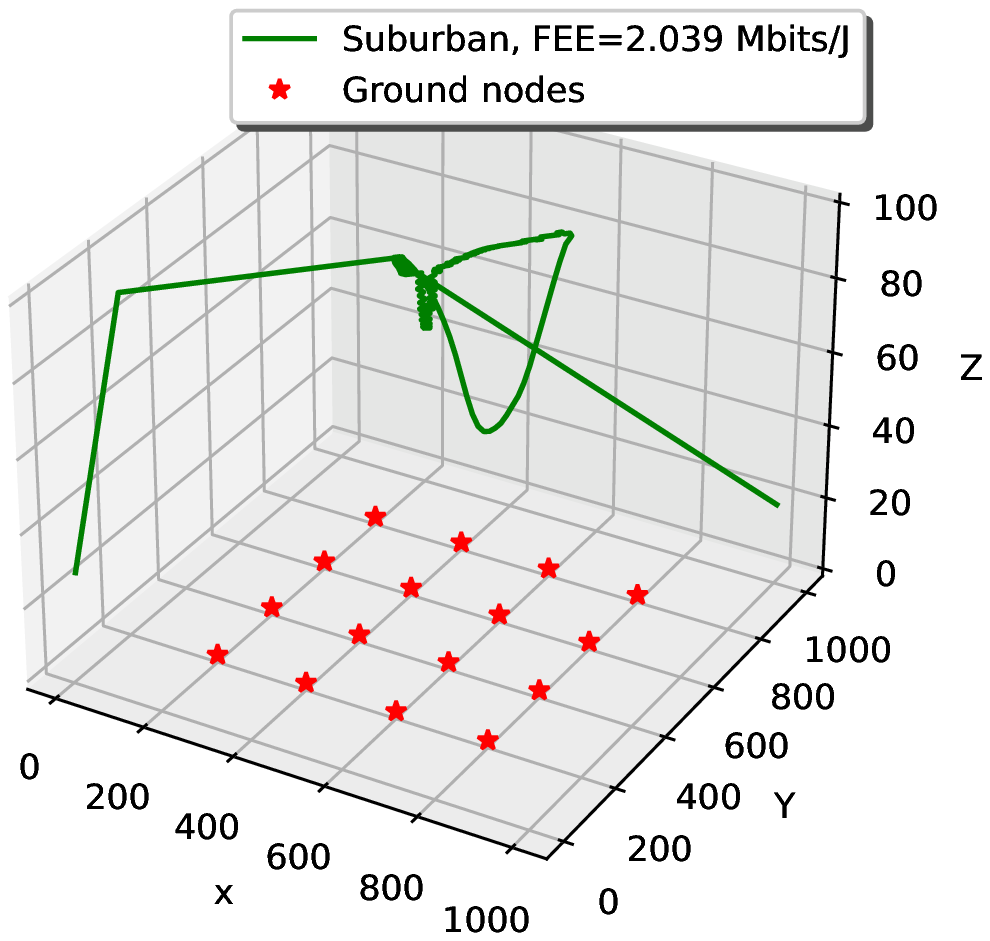}
         \caption{Trajectory.}
         \label{suburban}
     \end{subfigure}
     \hfill
     \begin{subfigure}[b]{0.49\columnwidth}
         \centering
         \includegraphics[width=\columnwidth]{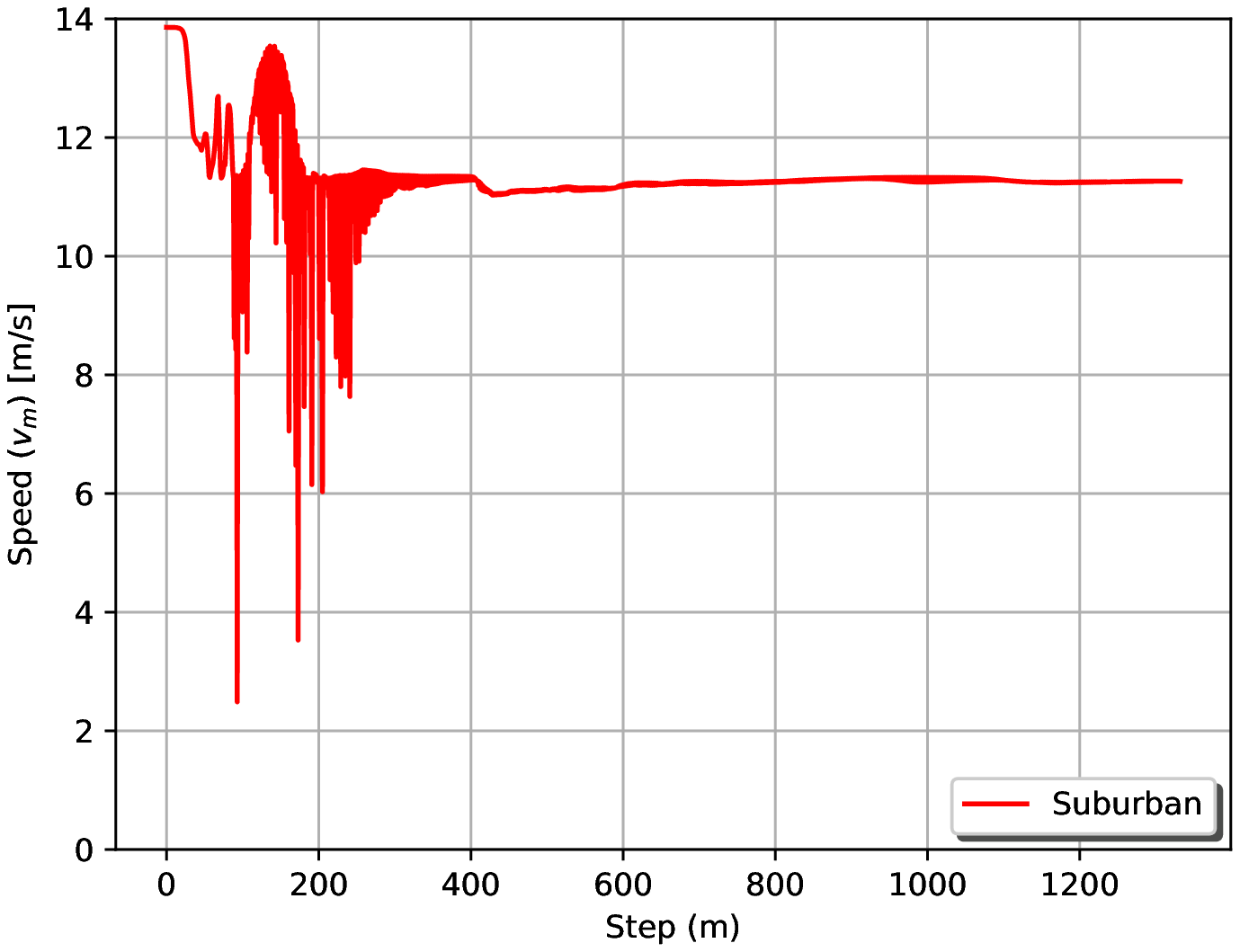}
         \caption{Speed.}
         \label{speedsuburban}
     \end{subfigure}
     \begin{subfigure}[b]{0.49\columnwidth}
         \centering
         \includegraphics[width=\columnwidth]{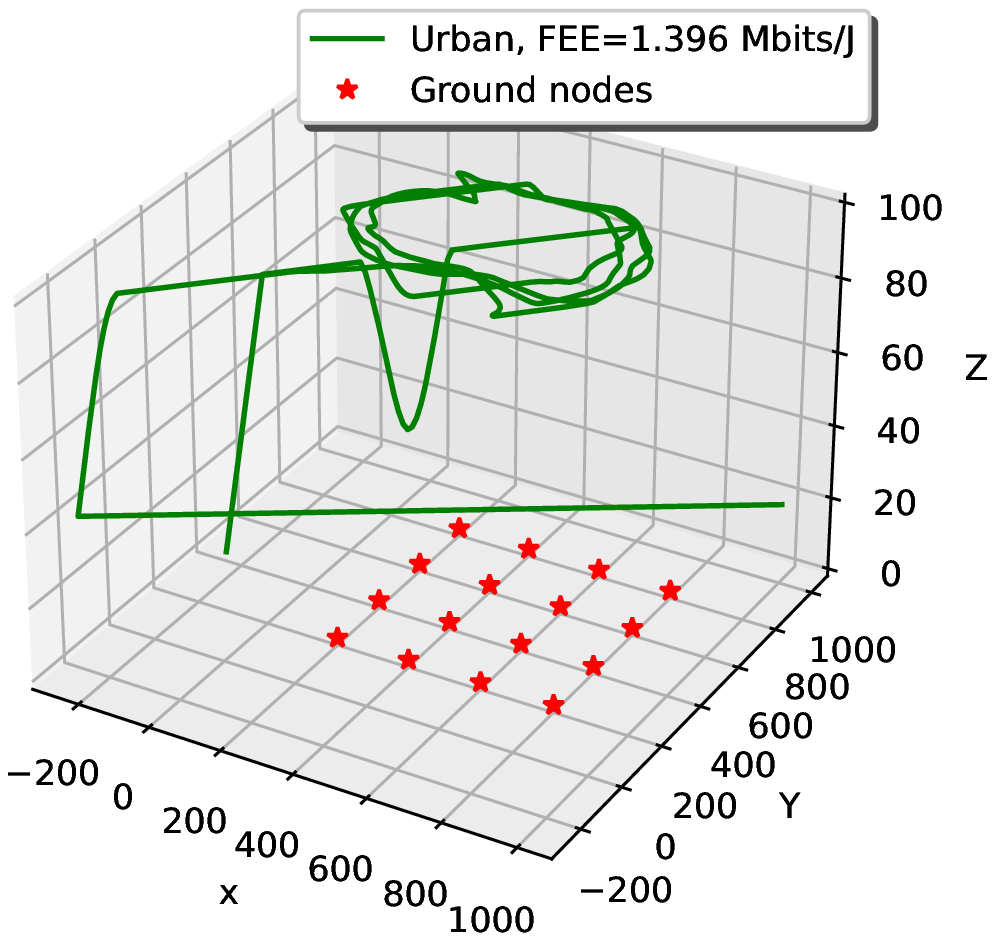}
         \caption{Trajectory.}
         \label{urban}
     \end{subfigure}
     \begin{subfigure}[b]{0.49\columnwidth}
         \centering
         \includegraphics[width=\columnwidth]{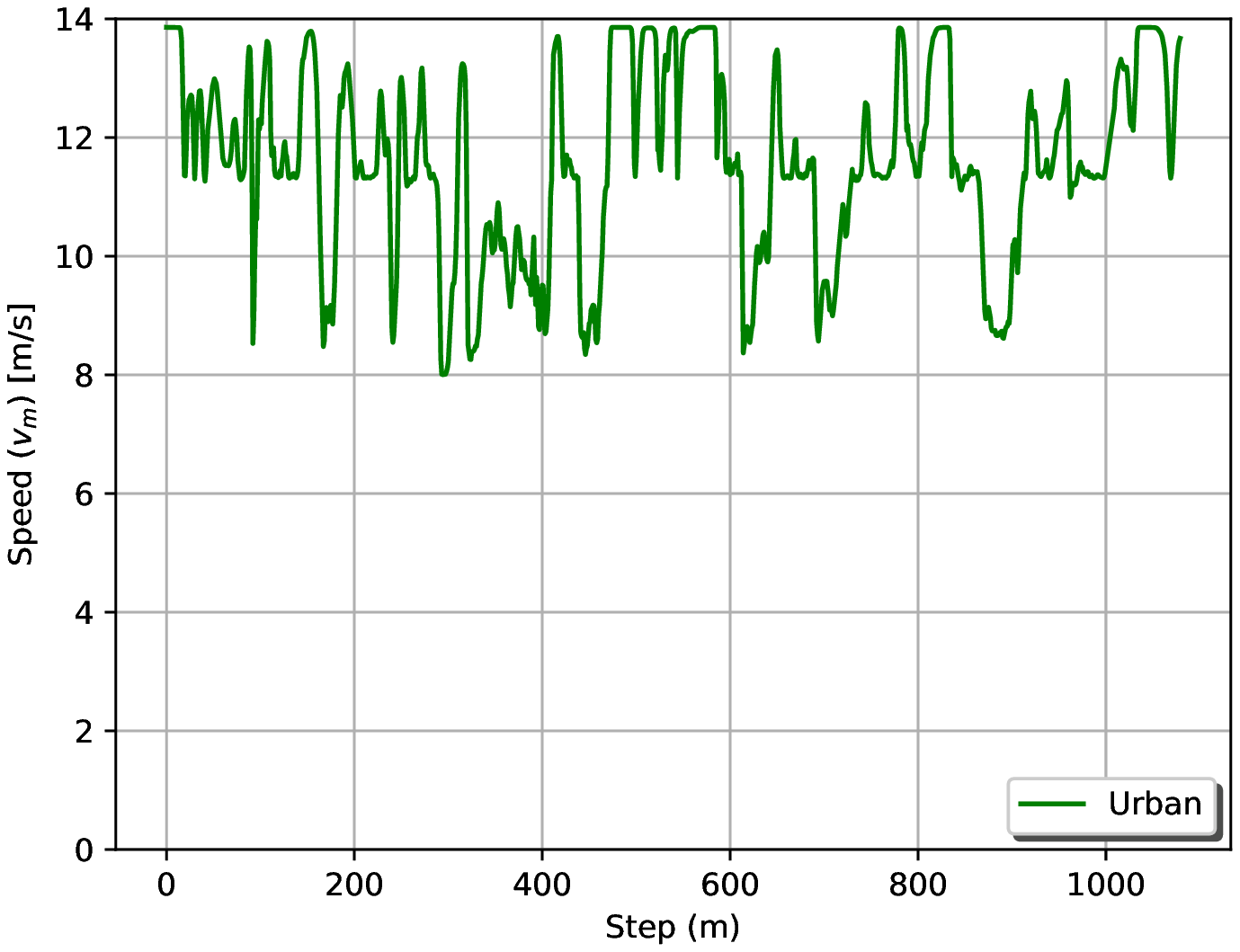}
         \caption{Speed.}
         \label{speedurban}
     \end{subfigure}
     \hfill
     \begin{subfigure}[b]{0.49\columnwidth}
         \centering
         \includegraphics[width=\columnwidth]{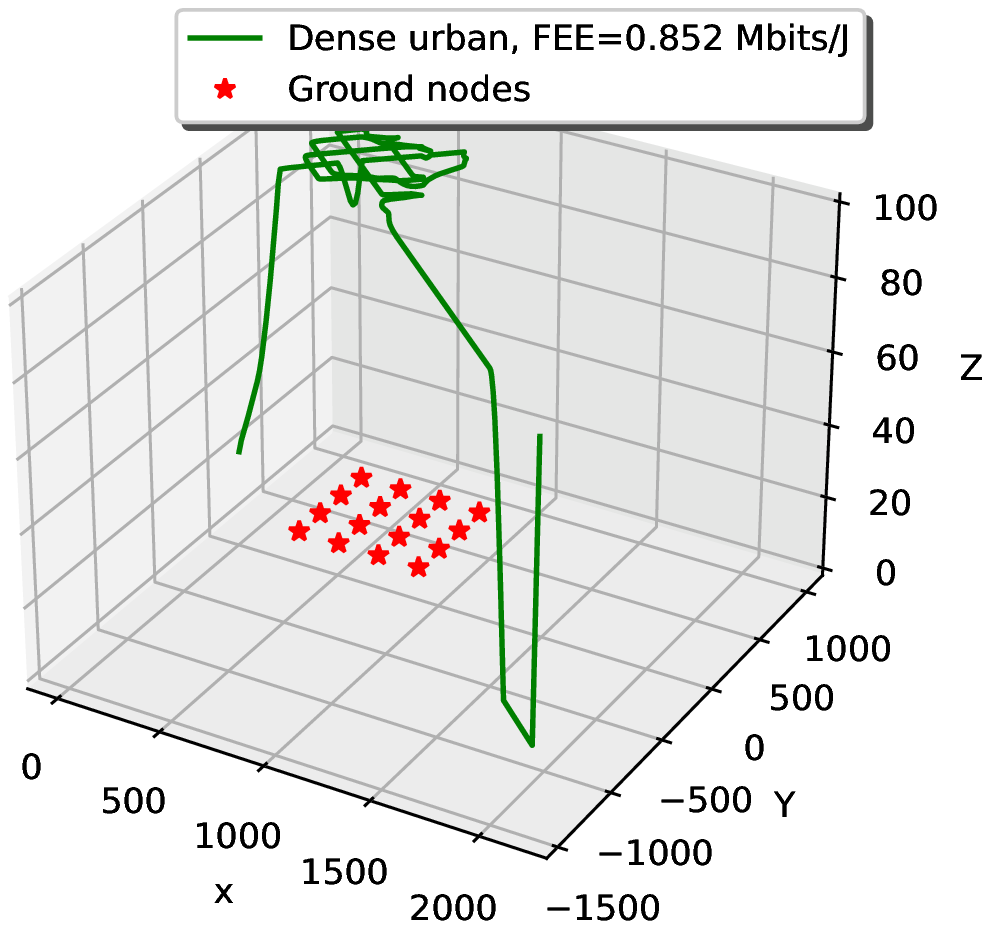}
         \caption{Trajectory.}
         \label{dense_urban}
     \end{subfigure}
     \begin{subfigure}[b]{0.49\columnwidth}
         \centering
         \includegraphics[width=\columnwidth]{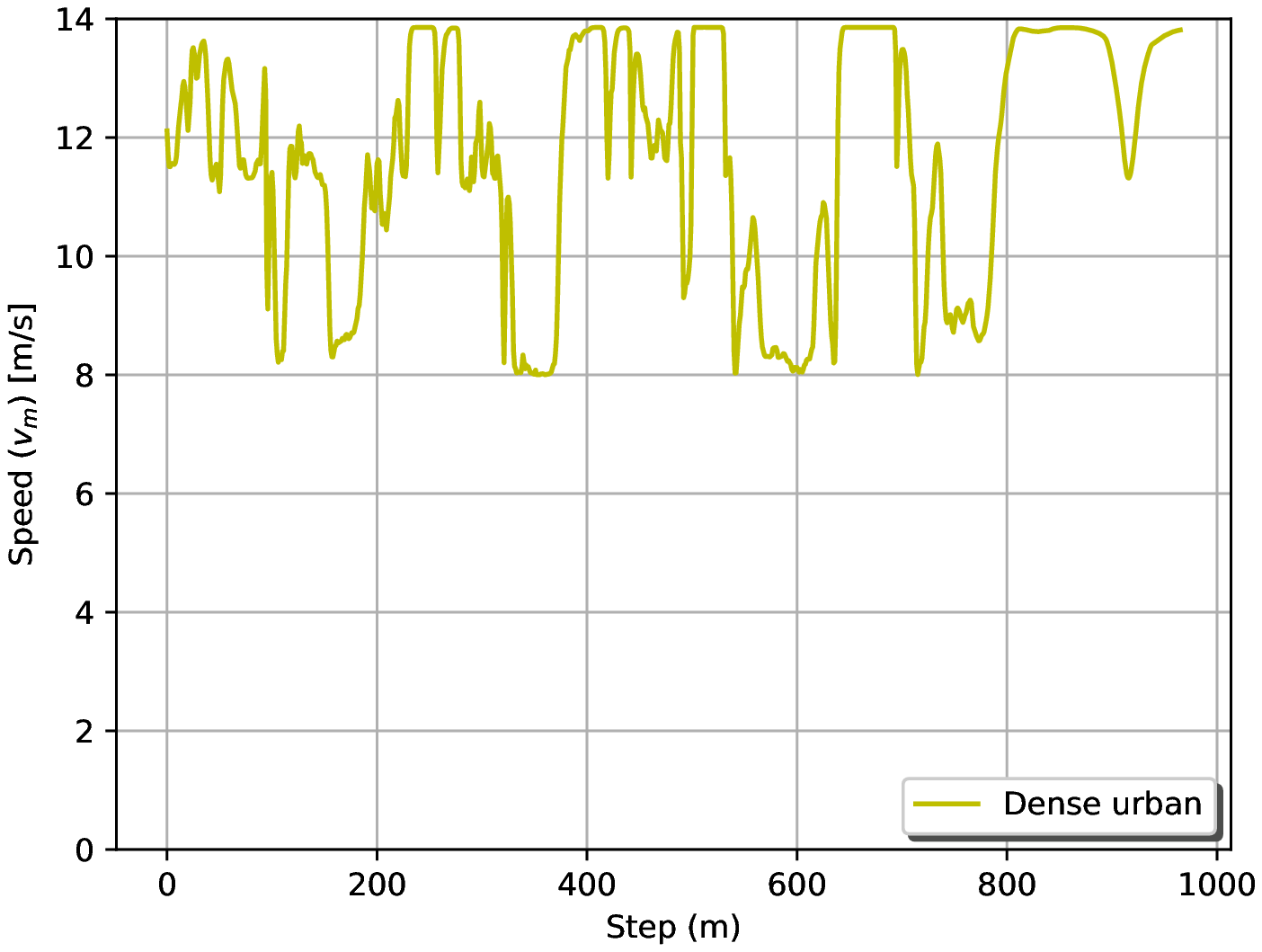}
         \caption{Speed.}
         \label{speeddurban}
     \end{subfigure}
        \caption{Sample PAP trajectories obtained using the trained actor network.}
        \label{trajectories}
\end{figure}
\begin{figure*}{}
\centering
    \begin{subfigure}[b]{0.8\columnwidth}
    \centering
    \includegraphics[width=\columnwidth]{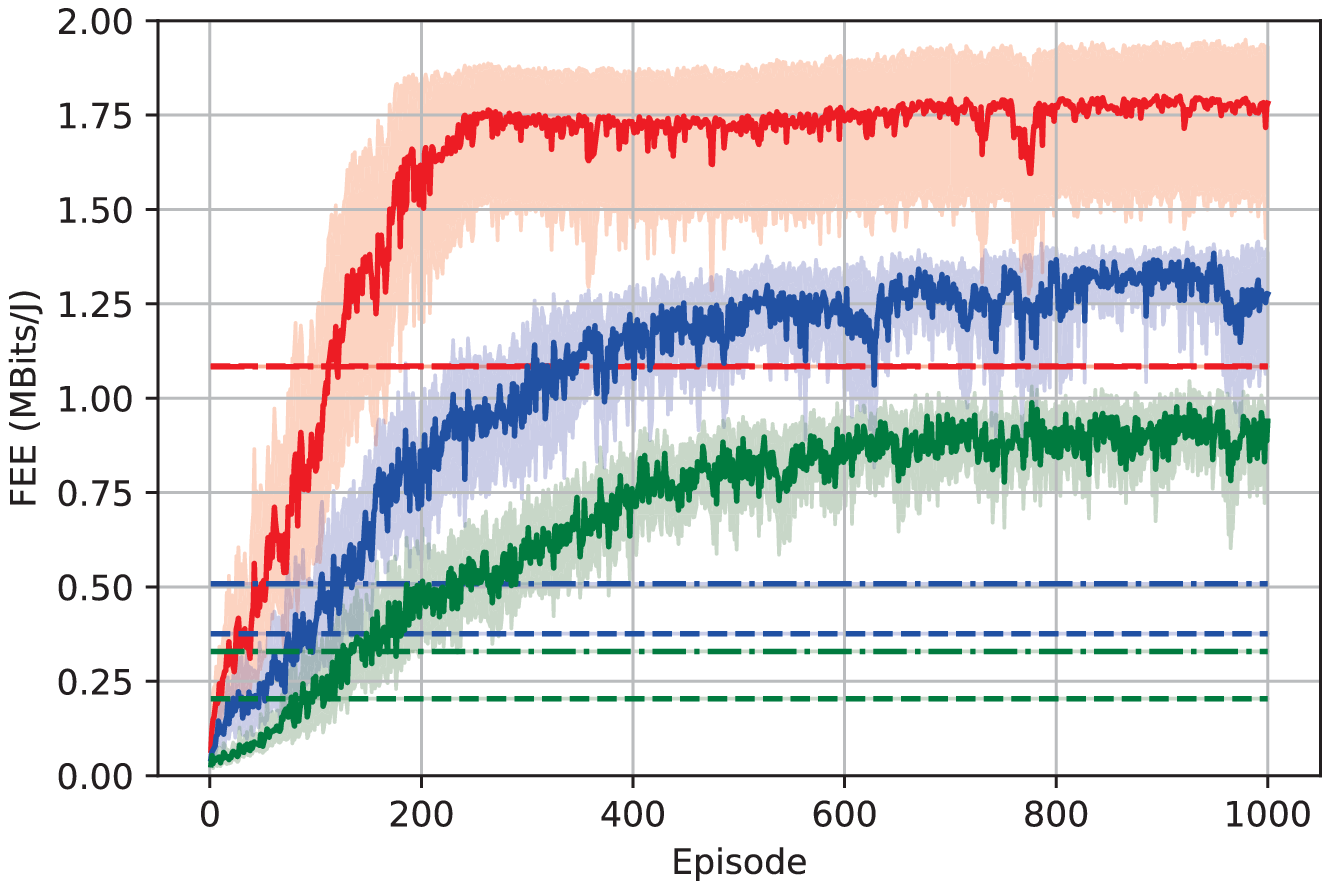}
    \caption{Fair energy efficiency.}\label{fee}
    \end{subfigure}
    \begin{subfigure}[b]{0.8\columnwidth}
    \centering
    \includegraphics[width=\columnwidth]{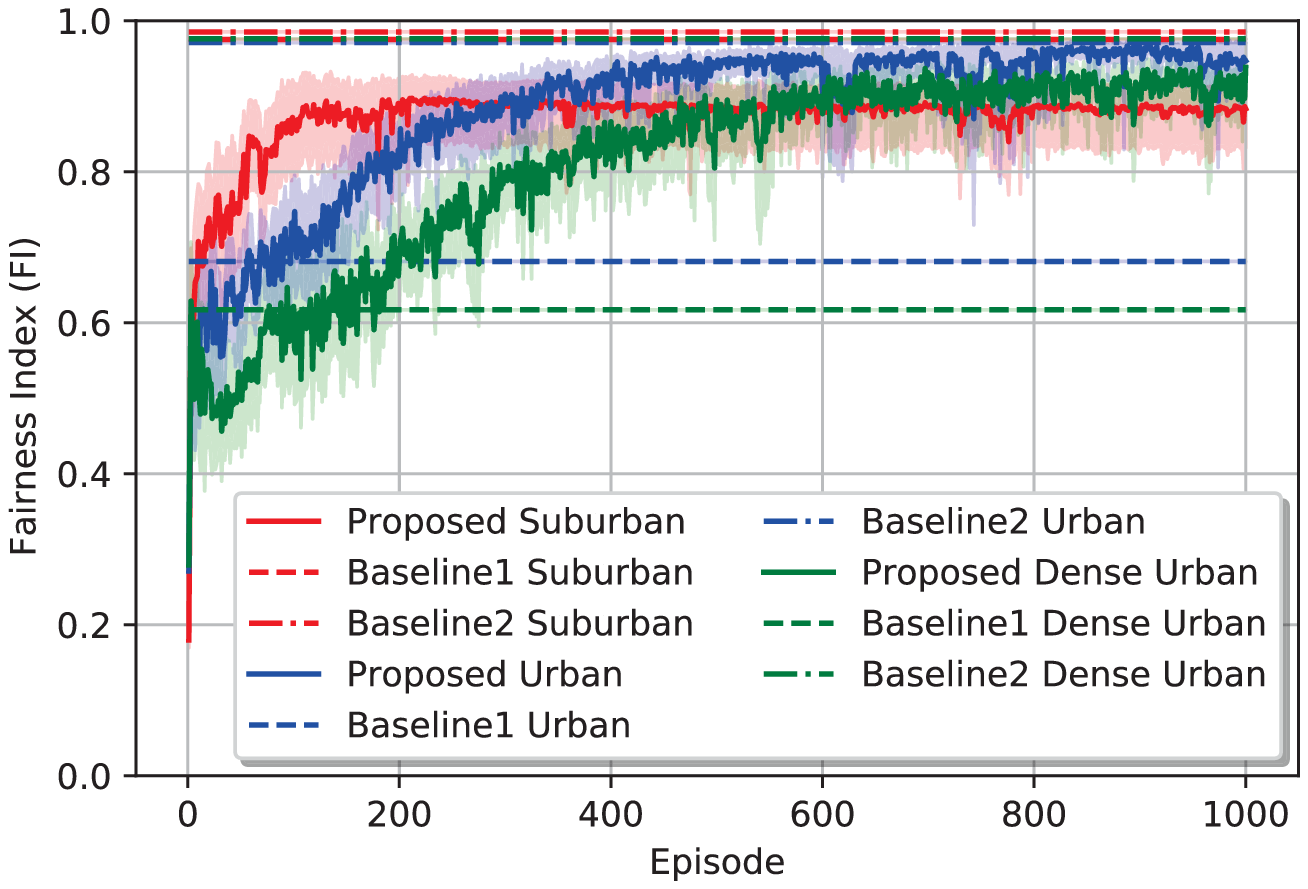}
    \caption{Fairness index.}\label{fi}
    \end{subfigure}
    \begin{subfigure}[b]{0.8\columnwidth}
    \centering
    \includegraphics[width=\columnwidth]{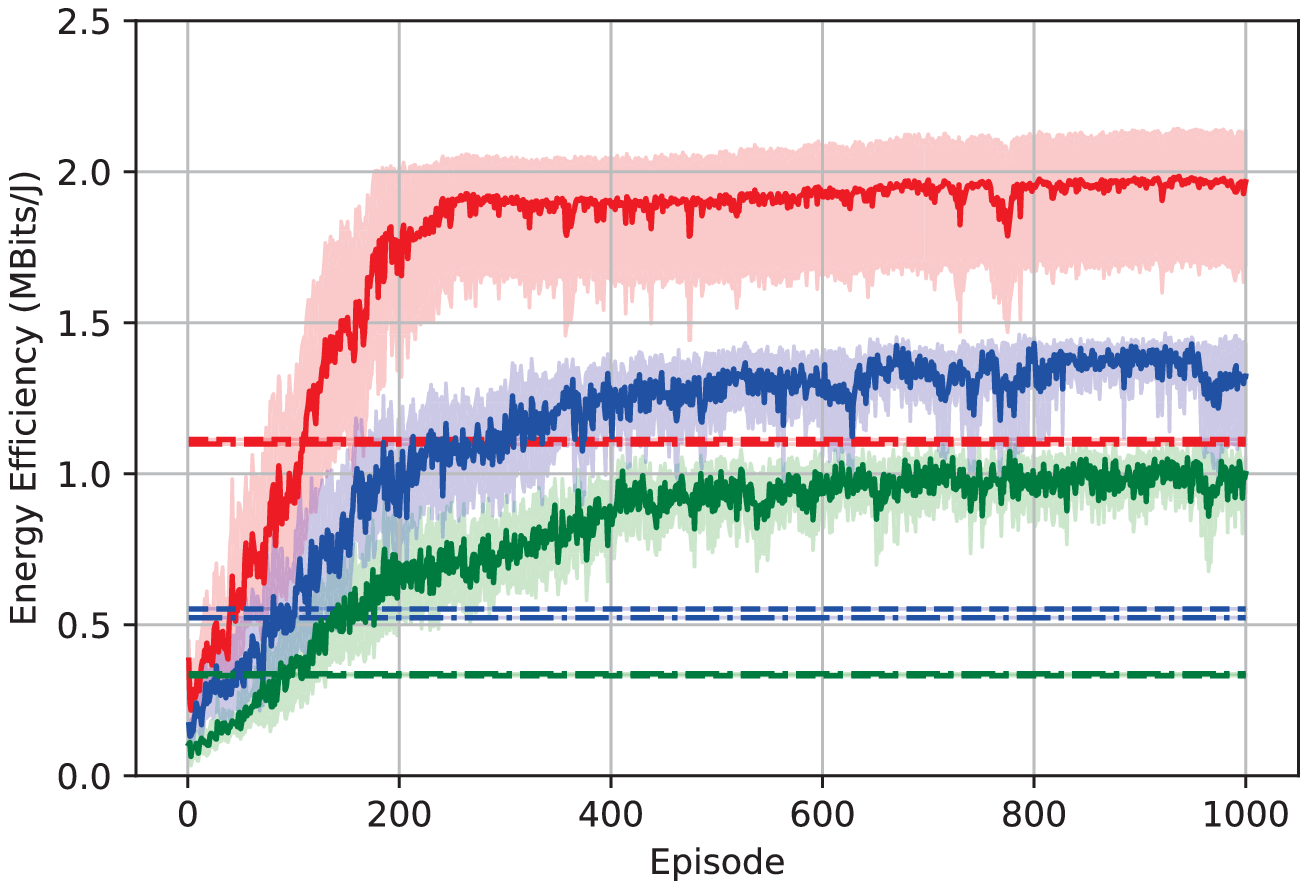}
    \caption{Energy-efficiency.}\label{ee}
    \end{subfigure}
    \begin{subfigure}[b]{0.8\columnwidth}
    \centering
    \includegraphics[width=\columnwidth]{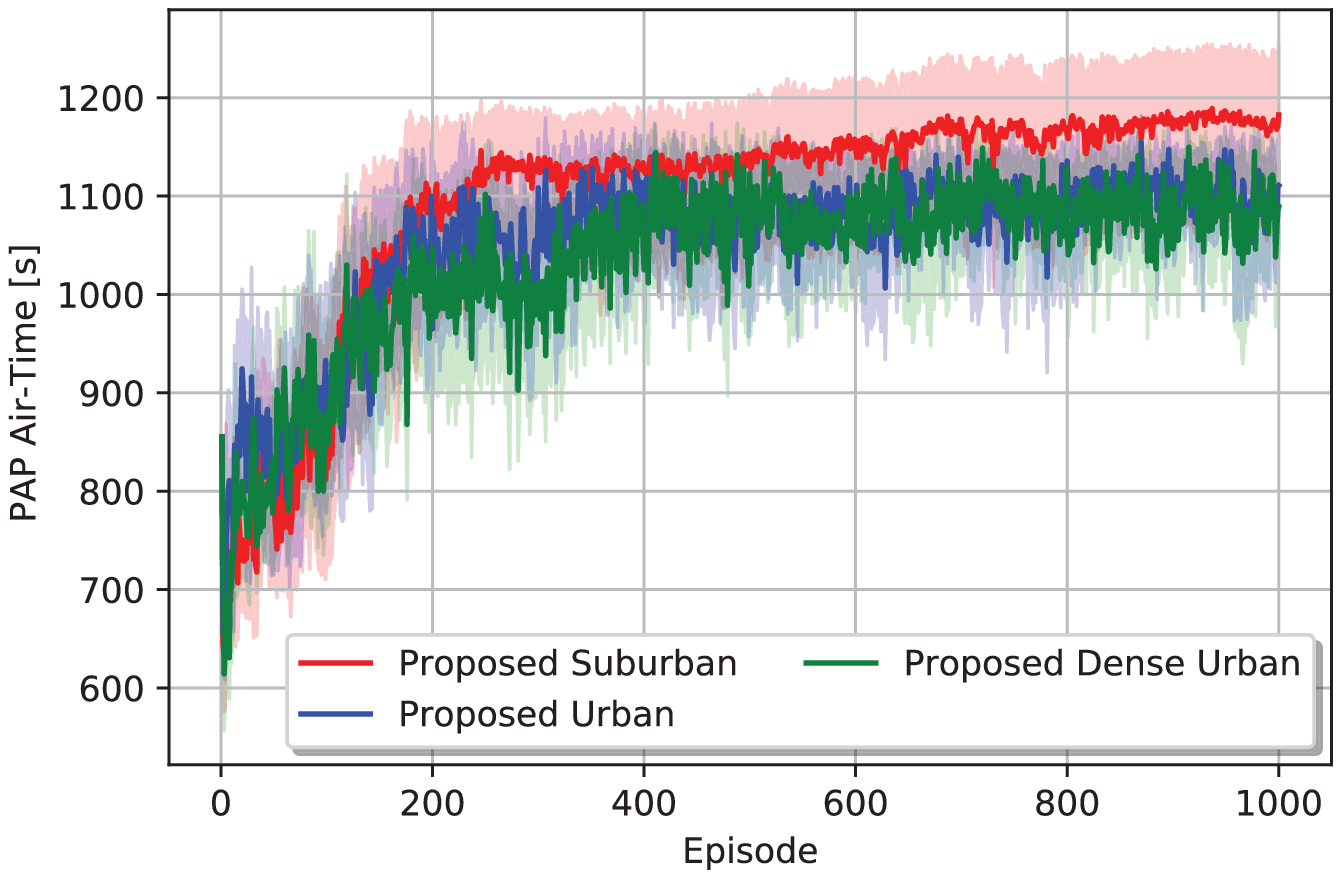}
    \caption{PAP Airtime. }\label{step}
    \end{subfigure}
    \caption{FEE, FI, EE, and PAP air-time improvements as the training progresses. The training procedure is repeated for 16 different random seeds. The shadow regions around the plots show a 95\% confidence interval across randomized repetitions. }
\end{figure*}
Finally, we perform two types of RL training to showcase the effectiveness of our approach in different scenarios. The first (offline) approach assumes the same fixed user positions for both the training and testing part. This analysis is done to evaluate the capability of the actor network to solve problem where all users are uniformly spaced along the x and y directions, as shown in Fig. \ref{baseline}. This fixed arrangement has very sparse GNs which makes it difficult for the FI problem. The second approach considers random positions, such as in a point process, where the x and y coordinates of each GN are uniformly distributed for each training episode. In this approach the testing is done on a set of random GN arrangements that the agent has not used for training before, which also includes the fixed uniform positions. This is a slightly easier problem when solving fairness problems such as FI due to the likelihood of users to cluster and/or disperse due to the entropy of the system.
\subsection{Fixed Uniform User Positions (Offline RL)}
The resulting trajectories following the training process of the actor network for the first arrangement are given in Fig. \ref{trajectories}. We notice that in the case of suburban deployments, the TD3 DRL method behaves very similarly to the stop-and-hover baseline deployment. This is due to the very high likelihood of having a LoS with all users and gives less relevance to the position of the PAP. However, opposite to the baseline case, the DRL method performs occasional repositioning to maintain better FEE. Finally, the TD3-DRL implementation keeps the PAP in constant movement with speeds around its most energy efficient velocity. We note that the most energy efficient velocity varies with the aerodynamics of the specific UAV and can thus be different with UAVs from different manufacturers.
Opposed to the suburban scenario, in Fig. \ref{trajectories} we can see in the subplots c) and e) that the PAP maintains much more dynamic movements for the urban and dense urban scenarios respectively. \color{black}This is a superior approach to the stop-and-hover one, due to the short bursts of better LoS connectivity when the PAP travels above each GN. However, these bursts need to be balanced over the longer period of service and thus the PAP is always kept on the move. \color{black}Finally, it is noticeable that in these two scenarios, the PAP flies off to a position that is far from the center of the area of service. This is significant for keeping the nodes equally serviced, and thus have improved FEE, in the more challenging propagation environments.

Figs. \ref{fee}-\ref{step} show the respective improvement in the fair energy efficiency, fairness index, energy efficiency, and PAP air time as the training progresses. As seen in the figures, initially, the agent tries random trajectories to explore the state-action space causing relatively shorter episodes with low FEE, FI, and EE values. Later, this experience helps the ML model to converge to a better policy that improves the optimization metric. \color{black}Moreover, the testing values (the values after 1000 episodes) using the proposed algorithm outperform the baseline scenarios in all the considered environments, as given in Table \ref{improvement}.\color{black}
\begin{table}[]
\centering
\caption{Improvement with respect to the \color{black}baselines\color{black}.}
\resizebox{0.9\columnwidth}{!}{%
\begin{tabular}{|l|l|l|l|l|l|l|}
\hline
                             &     & Proposed & Baseline1 & Improvement& \color{black}Baseline2& \color{black}Improvement\\ \hline
\multirow{3}{*}{Suburban}    & FEE & 2.039    & 1.085 & 87.93 \%  &\color{black}1.083 &\color{black} 88.27\% \\ \cline{2-7} 
                             & FI  & 0.896    & 0.975 & -8.1 \% &\color{black}0.985 &\color{black} -9.03\%\\ \cline{2-7} 
                             & EE  & 2.276    & 1.113 & 104.49\%  &\color{black}1.098 &\color{black}107.28\% \\ \hline
\multirow{3}{*}{Urban}       & FEE  & 1.4      & 0.376 &272.34\% &\color{black} 0.508 &\color{black} 175.59\%\\ \cline{2-7} 
                             & FI  & 0.968    & 0.681  & 42.14 \% &\color{black}0.971&\color{black}-0.31\%\\ \cline{2-7} 
                             & EE   & 0.876    & 0.552&58.69 \% &\color{black}   0.523&\color{black}67.49\%\\ \hline
\multirow{3}{*}{Dense urban} & FEE & 0.853    & 0.204  & 318.13 \%  &\color{black} 0.329&\color{black}159.27\%\\ \cline{2-7} 
                             & FI  & 0.902    & 0.617  &46.19 \% & \color{black}0.976&\color{black} -7.58\%\\ \cline{2-7} 
                             & EE   & 0.946    & 0.331 &185.8 \% &\color{black} 0.337&\color{black}180.71\% \\ \hline
\end{tabular}%
}
\label{improvement}
\end{table}
\color{black}The maximum gain over Baseline1 is achieved when the PAP is deployed in a dense urban scenario. This is because the considered setup places a subset of GNs in the NLoS regime of the PAP's hovering point, thereby giving a low baseline FI value. The proposed algorithm improves the FI value by moving the PAP around the GNs, as shown in Fig. \ref{dense_urban}. The performance gain achieved by Baseline2 is competitive in urban and dense urban scenarios, with regards to Baseline1, because the corresponding trajectory improves the service fairness among all the GNs. Interestingly, the FEE performances of Baseline1 and Baseline 2 in a suburban scenario are comparable since all the GNs are in LoS with the PAP throughout the respective placement policy, giving high FI values. This leads us to the conclusion that the suburban environment is not challenging enough for the problem of trajectory with regards to the scale of our implementation. \color{black}. In all the trajectories, the PAP climbs to the maximum altitude after leaving the starting point and then follows a horizontal flight during the remaining endurance: as the altitude increases, the throughput between the PAP and a node increases due to an improved  LoS probability between them; furthermore, the PAP power consumption during a vertical flight is much higher compared to a horizontal flight as shown in Fig. \ref{pow_end}. Hence, flying horizontally at the maximum altitude increases the PAP air time as well as the number of bits transmitted to the GNs thereby improving the FEE value of the system.  Also, the speed plots of Fig. \ref{trajectories} show that the actor proposes to fly the PAP at the optimal flying speed that maximizes the air-time of the PAP.
\begin{figure}{}
     \centering
     \begin{subfigure}[b]{\columnwidth}
         \centering
         \includegraphics[width=\columnwidth]{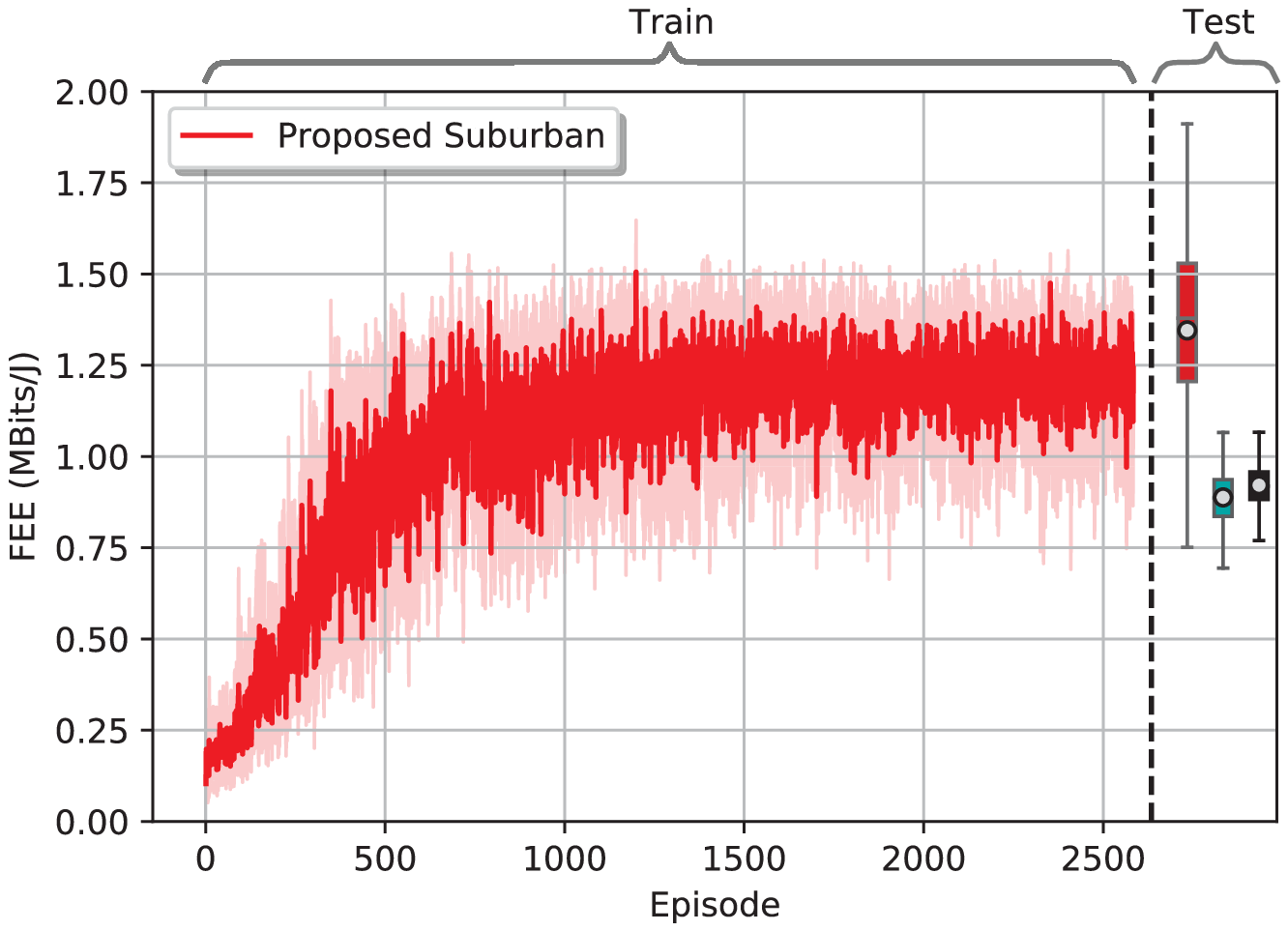}
         \caption{FEE Suburban}
         \label{suburban_fee}
     \end{subfigure}
     \begin{subfigure}[b]{\columnwidth}
         \centering
         \includegraphics[width=\columnwidth]{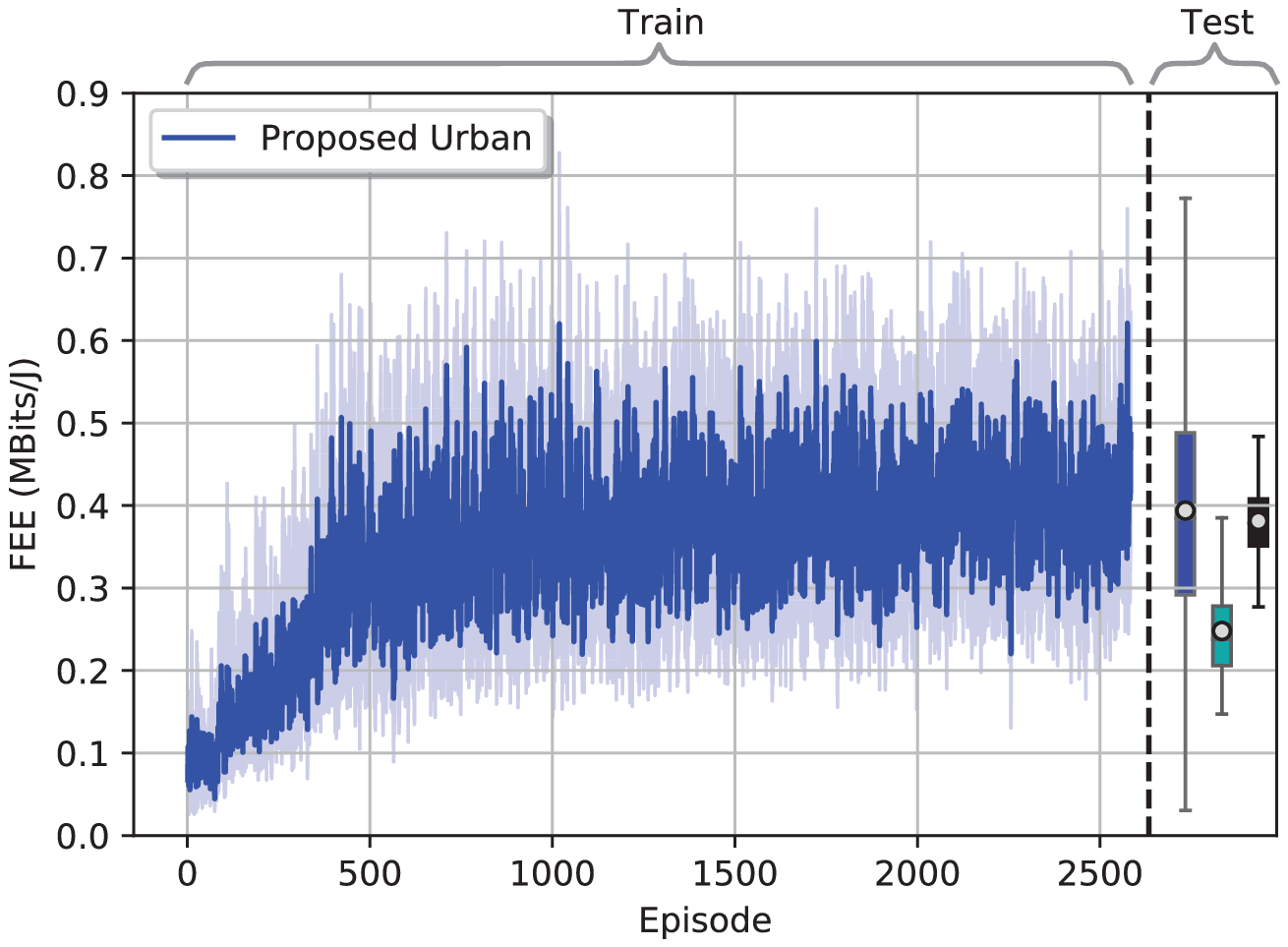}
         \caption{FEE Urban}
         \label{urban_fee}
     \end{subfigure}
     \begin{subfigure}[b]{\columnwidth}
         \centering
         \includegraphics[width=\columnwidth]{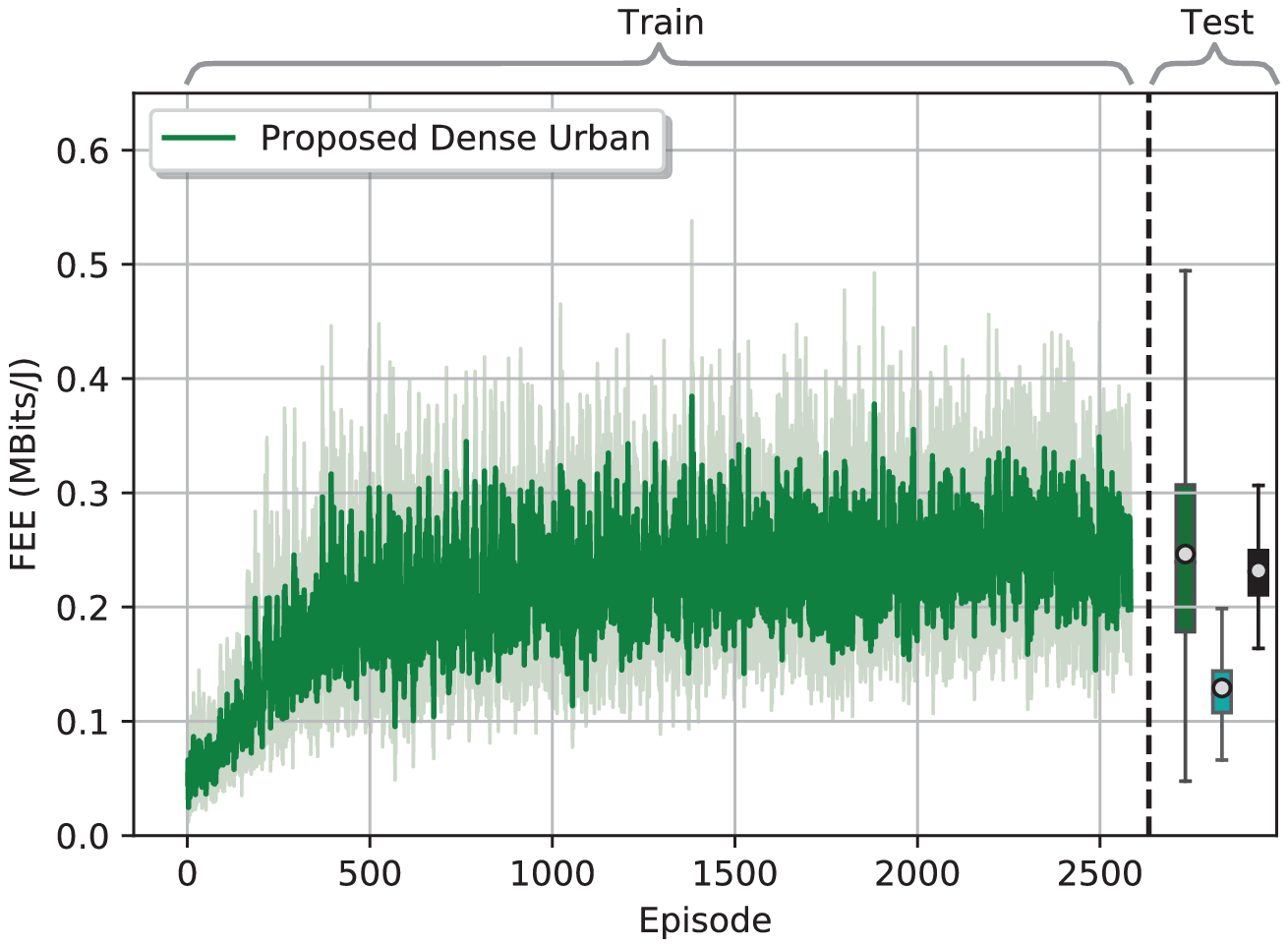}
         \caption{FEE Dense urban}
         \label{dense_urban_fee}
     \end{subfigure}
        \caption{Training and testing FEE plots. Circles inside the box plots represent the average test FEE values. \color{black}The second and third box plots represent Baselines 1 and 2, respectively.\color{black}}
        \label{train_test}
\end{figure}
\subsection{Randomized Uniform User Positions (Online RL)}
In this section, we describe the method adopted from \cite{mateus} to generalize the training so that the trained actor network performs well for any set of user positions. The system is trained for a fixed number of episodes $N_{\mathrm{train}}$, where the x and y coordinates of each GN are uniformly distributed for each training episode. After every 10 training episodes, we evaluate the actor network on a total of $N_{\mathrm{eval}}$ evaluation episodes with disabled learning to assess the current performance
of the actor network. We repeat this training procedure for $N_{\mathrm{seed}}$ times, each with a different random seed. The average FEE value after each evaluation phase is used as a metric to select the best-performing parameter ($\theta$) of the actor network ($\mu_\theta$). At the end of the training procedure, we further
evaluate the learned policy by assessing its performance in
$N_{\mathrm{test}}$ episodes, each with a different placement of GNs that the agent has not seen during the entire training process. Moreover, this test phase happens without exploration and learning. 

Fig. \ref{train_test} shows the training and testing performances for suburban, urban, and dense urban scenarios with $N_{\mathrm{train}}=1000$, $N_{\mathrm{eval}}=16$, and $N_{\mathrm{seed}}=8$. The shadow regions around the plots show a 95\% confidence interval across randomized repetitions. In all the scenarios, the FEE value improves with the training as seen in Fig. \ref{fee}. The mean and median FEE values obtained after testing the learned policy over $N_{\mathrm{test}}=512$ episodes outperform the mean baseline performances. As observed previously, the maximum and minimum performance gains are observed in dense urban and suburban scenarios, respectively. Additionally, the test performances (with learning disabled) are comparable with the performances at the end of the training phase. Thus the learned policy can be used to design an energy-efficient 3D trajectory for a PAP deployed to serve any given distribution of the GNs while guaranteeing per-user service fairness. 
\section{Conclusion}
\label{conclusion}
In this paper, we considered a UAV in the role of a portable access point (PAP) that aims to maximize the novel fairness-based energy efficiency metric, fair energy efficiency (FEE). Optimizing the energy-efficiency of PAPs is important but this should not come at the expense of service fairness. The method we propose here strikes a good balance between both. Moreover, we defined a pragmatic non-linear discharge behavior of the PAP battery, as the Peukert effect. As the first work to investigate the Peukert effect in PAP 3D trajectory optimization for wireless IoT services, we initially investigated the impact of the non-linearity of the energy storage. As such, we deducted that neglecting the Peukert effect overestimates the PAP air time which could force the PAP to perform an early landing. Given the non-convex FEE maximization problem with non-tractable constraints we proposed an adapted implementation of a twin delayed deep deterministic policy gradient deep reinforcement learning (TD3-DRL) framework. The optimal solutions provided by TD3-DRL varied by the properties of the propagation environment. The improvements of using the TD3-DRL in suburban scenarios are moderate with a gain  up to 80\% in suburban over the baseline scenarios and around 200\% and 300\% in the urban-and dense urban scenarios respectively. Finally, we generalize the network to any set of GN positions. Thus, we can summarize, that our TD3-DRL implementation provides a robust solution for PAP trajectory optimization in both strongly LoS and strongly NLoS environments.

\bibliographystyle{IEEEtran}
\bibliography{./main.bbl}

\end{document}